\documentclass[epj]{svjour}
\usepackage{latexsym}
\usepackage{graphics}
\usepackage[latin1]{inputenc}
\usepackage{rotating}
\begin{document}
\renewcommand{\textfraction}{0.00000000001}
\renewcommand{\floatpagefraction}{1.0}
\title{Quasi-free photoproduction of {\boldmath{$\eta$}}-mesons off  
{\boldmath{$^3$}}He nuclei}
\author{
  L.~Witthauer\inst{1}, 
  D.~Werthm\"uller\inst{1},  
  I. Keshelashvili\inst{1},
  P.~Aguar-Bartolom${\acute{\rm e}}$$^{2}$,      
  J.~Ahrens\inst{2},
  J.R.M.~Annand\inst{3},
  H.J.~Arends\inst{2},
  K.~Bantawa\inst{4}, 
  R.~Beck\inst{2,5},
  V.~Bekrenev\inst{6},
  A.~Braghieri\inst{7},
  D.~Branford\inst{8},
  W.J.~Briscoe\inst{9},
  J.~Brudvik\inst{10},
  S.~Cherepnya\inst{11},
  B.~Demissie\inst{9},
  M.~Dieterle\inst{1}, 
  E.J.~Downie\inst{2,3,9},
  P.~Drexler\inst{12},
  L.V.~Fil'kov\inst{11},
  A.~Fix\inst{13},
  D.I.~Glazier\inst{8},
  D.~Hamilton\inst{3},
  E.~Heid\inst{2},
  D.~Hornidge\inst{14},
  D.~Howdle\inst{3},
  G.M.~Huber\inst{15},
  I. Jaegle\inst{1},
  O.~Jahn\inst{2},
  T.C.~Jude$^{8}$, 
  A.~K{\"a}ser\inst{1},   
  V.L.~Kashevarov\inst{2,11},
  R.~Kondratiev\inst{16},
  M.~Korolija\inst{17},
  S.P.~Kruglov\inst{6}, 
  B.~Krusche\inst{1},     
  A.~Kulbardis\inst{6},
  V.~Lisin\inst{16},
  K.~Livingston\inst{3},
  I.J.D.~MacGregor\inst{3},
  Y. Maghrbi\inst{1},
  J.~Mancell\inst{3},   
  D.M.~Manley\inst{4},
  Z.~Marinides\inst{9},  
  M.~Martinez\inst{2},
  J.C.~McGeorge\inst{3},
  E.~McNicoll$^{3}$,  
  V.~Metag\inst{12},
  D.G.~Middleton\inst{14},
  A.~Mushkarenkov\inst{7},  
  B.M.K.~Nefkens\inst{10},
  A.~Nikolaev\inst{2,5},
  R.~Novotny\inst{12},
  M.~Oberle\inst{1},  
  M.~Ostrick\inst{2},
  B.~Oussena\inst{2,9},   
  P.~Pedroni\inst{7},
  F.~Pheron\inst{1},
  A.~Polonski\inst{16},
  S.~Prakhov\inst{2,9,10},
  J.~Robinson\inst{3},
  G.~Rosner\inst{3},
  M.~Rost\inst{2},
  T.~Rostomyan\inst{1},
  S.~Schumann\inst{2,5},
  M.H.~Sikora$^{8}$,  
  D.~Sober\inst{18},
  A.~Starostin\inst{10},
  I.~Supek\inst{17},
  M.~Thiel\inst{2,12},
  A.~Thomas\inst{2},
  M.~Unverzagt\inst{2,5},
  D.P.~Watts\inst{8}
\newline(Crystal Ball/TAPS experiment at MAMI, the A2 Collaboration)
\mail{B. Krusche, Klingelbergstrasse 82, CH-4056 Basel, Switzerland,
\email{Bernd.Krusche@unibas.ch}}
}
\institute{Department of Physics, University of Basel, CH-4056 Basel, Switzerland
  \and Institut f\"ur Kernphysik, University of Mainz, D-55099 Mainz, Germany
  \and SUPA, School of Physics and Astronomy, University of Glasgow, Glasgow G12 8QQ, UK
  \and Kent State University, Kent, Ohio 44242, USA  
  \and Helmholtz-Institut f\"ur Strahlen- und Kernphysik, University of Bonn, D-53115 Bonn, Germany
  \and Petersburg Nuclear Physics Institute, RU-188300 Gatchina, Russia
  \and INFN Sezione di Pavia, I-27100 Pavia, Pavia, Italy
  \and SUPA, School of Physics, University of Edinburgh, Edinburgh EH9 3JZ, UK
  \and Institute for Nuclear Studies, The George Washington University, Washington, DC 20052, USA
  \and University of California Los Angeles, Los Angeles, California 90095-1547, USA
  \and Lebedev Physical Institute, RU-119991 Moscow, Russia
  \and II. Physikalisches Institut, University of Giessen, D-35392 Giessen, Germany
  \and Laboratory of Mathematical Physics, Tomsk Polytechnic University, Tomsk, Russia
  \and Mount Allison University, Sackville, New Brunswick E4L 1E6, Canada
  \and University of Regina, Regina, SK S4S-0A2 Canada  
  \and Institute for Nuclear Research, RU-125047 Moscow, Russia
  \and Rudjer Boskovic Institute, HR-10000 Zagreb, Croatia
  \and The Catholic University of America, Washington, DC 20064, USA
}
\authorrunning{L. Witthauer et al.}
\titlerunning{Quasi-free photoproduction of $\eta$-mesons off $^3$He}

\abstract{
Quasi-free photoproduction of $\eta$-mesons has been measured off nucleons bound in 
$^3$He nuclei for incident photon energies from the threshold region up to 1.4 GeV. 
The experiment was performed at the tagged photon facility 
of the Mainz MAMI accelerator with an almost $4\pi$ covering electromagnetic calorimeter, 
combining the TAPS and Crystal Ball detectors. The $\eta$-mesons were detected in
coincidence with the recoil nucleons. This allowed a comparison of the production 
cross section off quasi-free protons and quasi-free neutrons and a full kinematic
reconstruction of the final state, eliminating effects from nuclear Fermi motion.     
In the $S_{11}$(1535) resonance peak, the data agree with the neutron/proton cross 
section ratio extracted from measurements with deuteron targets. More importantly,
the prominent structure observed in photoproduction off quasi-free neutrons bound in the
deuteron is also clearly observed. Its parameters (width, strength) are consistent with 
the expectations from the deuteron results. On an absolute scale the cross sections for both
quasi-free protons and neutrons are suppressed with respect to the deuteron target pointing
to significant nuclear final state interaction effects.
\PACS{
      {13.60.Le}{Meson production}   \and
      {14.20.Gk}{Baryon resonances with S=0} \and
      {25.20.Lj}{Photoproduction reactions}
            } 
} 
\maketitle

\section{Introduction}
\label{sec:intro}
Currently the only practical method to investigate the electromagnetic excitation 
spectrum of the neutron is photoproduction of mesons off nucleons bound in light nuclei.
Photoproduction of mesons off the free proton has been (and still is) intensively 
investigated with the measurement
of differential cross sections and single and double polarization observables for many
different final states (pseudoscalar mesons, vector mesons, meson pairs). The aim is 
to establish a reliable electromagnetic excitation scheme of the nucleon; however, since 
the electromagnetic interaction is isospin dependent, measurements with neutron targets are
mandatory for a complete picture. Programs to measure such reactions off neutrons 
bound in the deuteron are currently under way at several laboratories (see
\cite{Krusche_11} for an overview). Such measurements are complicated in comparison to 
those with a free proton target by three factors. The first is of technical nature, 
the indispensable detection of recoil nucleons in coincidence with the produced
mesons complicates the experiments. The detection of recoil neutrons 
lowers significantly the overall detection efficiency and the determination 
of the latter introduces an additional systematic uncertainty. Fermi motion of the bound
nucleons smears out all structures in the measured reaction cross section. It can significantly
modify angular distributions, in particular close to reaction thresholds, and thus must be
carefully considered when results are compared to model predictions for the free neutron
target. Finally, additional nuclear effects like meson-nucleon or nucleon-nucleon 
final-state interactions (FSI) may modify the measured cross sections with respect to the 
expectation from the elementary reactions off free nucleons. For neutrons bound in the 
deuteron, such effects have been studied in detail for different reaction channels \cite{Krusche_11}. 
In many cases, cross sections and other observables measured off free and quasi-free protons 
are in good agreement (examples are 
\cite{Ajaka_07,Fantini_08,DiSalvo_09,Jaegle_11b,Jaegle_11,Oberle_13} 
for $\eta$, $\pi^0$, $\eta'$, and $\pi^0\pi^0$ photoproduction), which is the basis for the 
extraction of such data for neutron targets from quasi-free deuteron data.     

Other light nuclei are less frequently used for such measurements, although two of them have
certain advantages. Tritium ($^3$H) nuclei offer the most favorable neutron/proton ratio,
but the difficulties in handling them usually disfavor their use. Measurements of observables 
involving the polarization degree of freedom of the initial nucleon may take advantage of the 
spin structure of $^3$He nuclei. For the main component of their wave function the two protons 
are coupled with antiparallel spin so that the net spin of the nucleus is identical with the 
spin of the neutron. An experimental programme to measure the helicity decomposition of 
photon-induced reactions with polarized $^3$He gas targets and circularly polarized photon 
beams is underway at the MAMI facility \cite{Bartolome_13,Constanza_13}. 

A particularly 
interesting reaction is the photoproduction of $\eta$-mesons off the neutron. Previous 
measurements using deuterium targets revealed a narrow structure of unexplained nature in
the excitation function of $\gamma n\rightarrow n\eta$ 
\cite{Jaegle_11,Kuznetsov_07,Miyahara_07,Jaegle_08,Werthmueller_13}. The present experiment 
used $^3$He nuclei as a quasi-free neutron target. This allows the study of the structure 
in a different nuclear environment to test whether it behaves as expected due to the 
different momentum distributions of the nucleons in $^2$H and $^3$He nuclei and
to investigate the effects of FSI on $\eta$ photoproduction from $^3$He nuclei.
  
Coherent photoproduction of $\eta$-mesons from $^3$He has also been analyzed in a 
separate study to provide information on the formation of eta-mesic nuclei
\cite{Pheron_12}.   

\section{Resonance contributions to photoproduction of {\boldmath{$\eta$}}-mesons}

Photoproduction (also with virtual photons) of $\eta$-mesons off free protons has been 
studied in much detail. Angular distributions and the total cross section have been 
measured from threshold ($E_{\gamma}^{\rm thr}$ = 707.8 MeV) up to incident photon energies 
of $\approx$ 3 GeV by different experiments 
\cite{Krusche_95,Armstrong_99,Thompson_01,Renard_02,Dugger_02,Crede_05,Nakabayashi_06,Bartalini_07,Bartholomy_07,Denizli_07,Crede_09,Williams_09,Sumihama_09,McNicoll_10}.    

Photoproduction of $\eta$-mesons in the threshold region is characterized  by the 
dominance of the $S_{11}$(1535) resonance \cite{Krusche_97,Krusche_95}. Contributions 
from the close-by $D_{13}$(1520) are small. They have been identified via interference 
terms in the angular distributions \cite{Krusche_95} and in particular in the photon-beam
asymmetry $\Sigma$ \cite{Ajaka_98,Elsner_07}. An analysis in the framework of the 
`Eta-MAID' model \cite{Chiang_02} allowed the extraction of the tiny  $N\eta$ branching 
ratio (0.23$\pm$0.04 \%) \cite{PDG} of the $D_{13}$ resonance. At somewhat higher energies, 
for final-state invariant mass $W$ between 1.65 GeV and 1.75 GeV, the reaction 
is less well understood. Models agree 
on a contribution from the $S_{11}$(1650) resonance, which interferes with 
the $S_{11}$(1535) \cite{Chiang_02}. For the proton target the interference is clearly 
destructive, however, for the neutron the situation is unclear. The Particle Data Group (PDG)
\cite{PDG} quotes a negative value for the electromagnetic helicity coupling $A_{1/2}^n$ 
of this state for the neutron, while the recent analysis in the framework of the Bonn-Gatchina
(BnGn) coupled channel analysis \cite{Anisovich_13} finds a positive sign. The $D_{15}$(1675) 
has only a small electromagnetic coupling for the proton and its contribution is not well 
established. The $D_{13}$(1700) could also be present but its properties are poorly known. 
Model results for the $P_{11}$(1710) and the $P_{13}$(1720) are in conflict. As discussed 
in \cite{Elsner_07} in the `Eta-MAID' model \cite{Chiang_02} the $P_{11}$(1710) is more 
important than the $P_{13}$(1720). On the other hand, in the BnGn analysis \cite{Anisovich_05} 
the $P_{11}$ makes an almost negligible contribution while the $P_{13}$ is essential to 
describe the beam asymmetries. At even higher incident photon energies, more ambiguities exist.  

Quasi-free and coherent photoproduction of $\eta$-mesons off light nuclei 
($^2$H \cite{Krusche_95a,Hoffmann_97,Weiss_03,Weiss_01}, $^3$He \cite{Pfeiffer_04}, 
$^4$He \cite{Hejny_99,Hejny_02}) has so far been studied mainly in the excitation range 
of the $S_{11}$(1535) in order to pin down the isospin structure of the electromagnetic 
excitation of this state. The main result (see Ref. \cite{Krusche_03} for a summary) is a 
neutron/proton cross section ratio of $\approx$ 2/3 corresponding to a dominant isovector 
excitation of the $S_{11}$(1535) with $A_{1/2}^{IS}/A_{1/2}^{p}=$ 0.09$\pm$0.01, where 
$A_{1/2}^{p}$ is the helicity-1/2 coupling for the proton and $A_{1/2}^{IS}$ its
isoscalar component. At higher incident photon energies, models \cite{Chiang_02}
predicted a much larger contribution of the $D_{15}$(1675) state in the neutral channel, 
so that the neutron/proton cross section ratio should rise. However, the  
experimental finding \cite{Jaegle_11,Kuznetsov_07,Miyahara_07,Jaegle_08,Werthmueller_13} 
was the pronounced narrow structure in the $n\eta$-excitation function, which did 
not resemble the expected contribution from the $D_{15}$ state \cite{Jaegle_11}. 
Different scenarios for the nature of this structure have been discussed in the literature, 
among them coupled-channel effects from known nucleon resonances \cite{Shklyar_07,Shyam_08}, 
threshold effects from opening strangeness production \cite{Doering_10}, interference effects 
in the $S_{11}$-partial wave \cite{Anisovich_13,Anisovich_09}, but also intrinsically narrow excited 
nucleon states \cite{Anisovich_09,Arndt_04,Fix_07,Shrestha_12}. Until now no final 
conclusion has been drawn, but the results suggest \cite{Jaegle_11,Kuznetsov_07,Werthmueller_13}, 
that the structure is indeed very narrow (with an intrinsic width below 50~MeV).
Although most known nucleon resonances in this energy range are much broader, Shresta and 
Manley \cite{Shrestha_12} extracted with a recent multichannel partial wave analysis parameters 
of the $D_{13}$(1700) resonance which are similar to the structure observed in 
$\gamma n\rightarrow n\eta$. They reported a mass of 1665$\pm$3~MeV, a width of 56$\pm$8~MeV, 
and a photon coupling which is twice as large for the neutron than for the proton.
However, the decay branching ratio of this state into $\eta N$ is not known and according 
to the PDG \cite{PDG} it is at most in the 1\% range. 

\section{Experimental setup}
\label{sec:setup}
The measurements were performed at the MAMI accelerator in Mainz 
\cite{Herminghaus_83,Kaiser_08}. It delivered an electron beam of 1508~MeV energy 
with an intensity of 8~nA, which was used to generate bremsstrahlung photons in a 
radiator (copper foil of 10 $\mu$m thickness). The photons were tagged with the upgraded
Glasgow magnetic spectrometer \cite{Anthony_91,Hall_96,McGeorge_08} in the range from 
0.45~GeV to 1.4~GeV with a typical energy resolution of 4 MeV, defined by the geometrical 
width of the electron counters in the focal plane of the device. Liquid $^3$He was used
as target material. The target cell was a mylar cylinder of 3.0~cm diameter and 
5.08~cm length. At a temperature of 2.6~K a target surface density of 0.073 nuclei/barn 
was reached.  

\begin{figure}[thb]
\resizebox{0.50\textwidth}{!}{%
  \includegraphics{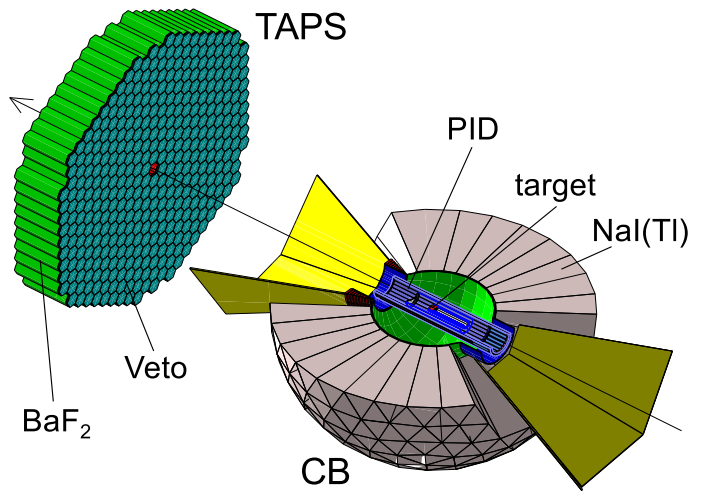}
}
\caption{Experimental setup of Crystal Ball (only bottom hemisphere shown)
with PID detector and TAPS forward wall.
}
\label{fig:setup}       
\end{figure}

Photons from the decay of the $\eta$-mesons and the recoil nucleons were detected with 
an electromagnetic calorimeter covering almost the full solid angle. A schematic drawing 
of the setup is shown in Fig.~\ref{fig:setup}.
It combines the Crystal Ball (CB) \cite{Starostin_01} with the TAPS detector 
\cite{Novotny_91,Gabler_94}. 
The CB is made of 672 NaI(Tl) crystals arranged in two 
hemispheres with the overall geometry of a sphere. The target was mounted in the center 
of this sphere, which covers the full azimuthal angle for polar angles from 20$^{\circ}$ 
to 160$^{\circ}$, corresponding to 93\% of the full solid angle. The forward angular 
range was covered by the TAPS detector configured as a hexagonal wall of 384 BaF$_2$ 
crystals. It was placed 1.457~m downstream from the target and covered polar angles 
between 5$^{\circ}$ and 21$^{\circ}$. For the identification of charged particles 
(recoil protons, charged pions) all modules of the TAPS detector had individual plastic 
scintillators (5~mm thickness) in front, from which also time and energy information was 
read (TAPS-CPV detector). Inside the CB a Particle Identification Detector (PID) 
\cite{Watts_04} was mounted around the target, which allowed identification of protons and 
charged pions via $E-\Delta E$ measurements. More details about the detectors are 
given in \cite{Schumann_10,Zehr_12}.

Since the experiment aimed mainly at a measurement of $\eta$-production using the 
$\eta\rightarrow 2\gamma$ decay, the main trigger condition required two hits in the 
calorimeter. For this purpose, the CB and the TAPS detector were subdivided into 
logical sectors: The TAPS detector into 6$\times$64 modules in a pizza-like geometry and 
the CB into 45 groups of 16 modules (the groups overlapping with the beam entrance and
exit holes had fewer modules). Events were accepted when detector modules in at least 
two logical sectors had signals above a threshold of 25 MeV and the analog energy-sum signal
from the CB was above 300 MeV. Events with two hits in TAPS and no hit in CB were thus 
not accepted, which is, however, irrelevant for $\eta\rightarrow  2\gamma$ decays. 
Due to the relatively large opening angle of the decay photons such events are rare. 
The analysis used only events for which in the course of the analysis these conditions 
were fulfilled by the 
$\eta$-decay photons. Events for which the trigger was only activated due to the energy 
deposition of the recoil nucleon were discarded in order to avoid systematic uncertainties
(the energy thresholds for the trigger were calibrated for photon showers, not for recoil 
nucleons). For accepted events the readout thresholds for the detector modules were 
set to 2 MeV for the CB-crystals, to 5 MeV for the TAPS crystals, to 250 keV for the 
TAPS-CPV, and to 300 keV for the elements of the PID.

\section{Data analysis}
\label{sec:ana}
Details of the calibration and analysis procedures for all detector components are given in 
\cite{Schumann_10,Zehr_12}. Here, we summarize only the most important steps of the present 
analysis.

\subsection{Particle and reaction identification} 
\label{ssec:iden}

\begin{figure}[htb]
\resizebox{0.50\textwidth}{!}{%
  \includegraphics{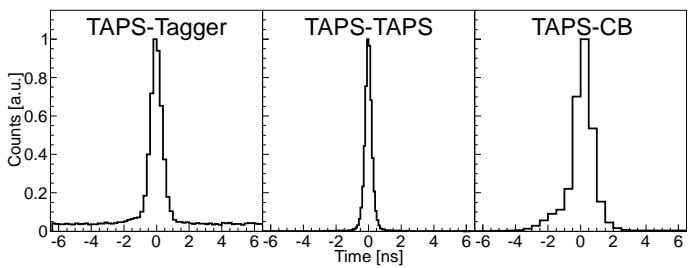}
}
\caption{Time coincidence spectra between photons in TAPS and electrons in the the tagger 
(left-hand side), between photon hits in TAPS (center), and between a photon hit in TAPS
and a photon hit in CB (right-hand side).
}
\label{fig:coinc}       
\end{figure}

The calorimeter was operated in coincidence with the tagging spectrometer. Background from 
random coincidences between Tagger and production detector was subtracted by a side-band 
analysis in the time-coincidence spectra. Typical timing spectra are shown in 
Fig. \ref{fig:coinc}. The fastest component of the setup was the TAPS detector with
a resolution of 0.5 ns (FWHM) for two photon hits in this calorimeter. The resolution
of TAPS relative to the Tagger was 0.8 ns and between TAPS and CB was 1.5 ns.

The analysis combined the particle identification capabilities of the experiment 
(charged-particle identification with TAPS-CPV and in the PID, pulse-shape analysis for TAPS,
time-of-flight (ToF) versus energy for TAPS, and $\Delta E - E$ with PID and CB) with the
reaction identification via invariant mass and missing mass spectra. The basis of the 
different measurements and the detector performance are discussed in more detail in 
\cite{Schumann_10,Zehr_12}.

The $\eta\rightarrow 2\gamma$ and the $\eta\rightarrow 3\pi^0\rightarrow 6\gamma$ 
decays of the mesons were analyzed in coincidence with protons ($\sigma_p$), in coincidence 
with neutrons ($\sigma_n$), and without any condition for recoil nucleons ($\sigma_{\rm incl}$ 
inclusive reaction, recoil nucleon may have been detected but was not required). The accepted 
event classes are summarized in Table~\ref{tab:events}. Events with additional hits were 
discarded in order to reduce background contributions, for example, from the photoproduction 
of $\eta\pi$-pairs (overlap of events originating from different incident photons
was negligible as can be seen in the TAPS-TAPS and TAPS-CB time spectra in Fig.~\ref{fig:coinc}). 
\begin{table}[h]
\begin{center}
\caption{Selected event classes for the cross sections $\sigma_p$, $\sigma_n$, and
$\sigma_{\rm incl}$ for the two $\eta$-decay branches. $n$ and $c$ mark neutral and charged hits 
in the calorimeter (distinguished by the response of the charged-particle detectors).
} 
\label{tab:events}       
\begin{tabular}{|c|c|c|c|}
\hline
& $\sigma_p$ & $\sigma_n$ & $\sigma_{\rm incl}$ \\
\hline
$\eta\rightarrow 2\gamma$ & 2$n$ \& 1$c$ & 3$n$ & 2$n$ or 3$n$ or (2$n$ \& 1$c$) \\
$\eta\rightarrow 6\gamma$ & 6$n$ \& 1$c$ & 7$n$ & 6$n$ or 7$n$ or (6$n$ \& 1$c$) \\
 \hline
\end{tabular}
\end{center}
\end{table}

\begin{figure}[thb]
\centerline{\resizebox{0.5\textwidth}{!}{%
  \includegraphics{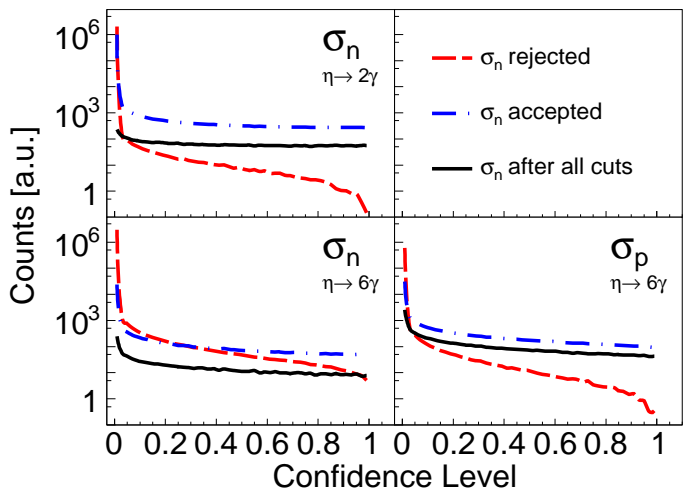}  
}}  
\caption{Confidence levels from $\chi^2$-tests. Top row: $n\eta\rightarrow n 2\gamma$
final state. Dashed (blue) histogram: `best' two-photon combinations, dashed (red) histogram: 
rejected combinations, solid (black) histogram: accepted after all further cuts.
Bottom row: $n\eta\rightarrow n6\gamma$ and $p\eta\rightarrow p6\gamma$ final states.
}
\label{fig:confl}       
\end{figure}

The first step of the analysis was thus the classification of detector hits as `charged' or
`neutral' using the information from the PID and TAPS-CPV. Subsequently, a pre-selection 
of $\eta$-candidates was done.
For tentative $\eta\rightarrow 2\gamma$ decays the $\chi^2$ defined by 
\begin{equation}
\chi^2 = \frac{(m( 2\gamma ) - m_{\eta})^2}{(\Delta m( 2\gamma ))^2}\;\;\;  
\label{eq:chi}
\end{equation} 
was calculated for all pairs of neutral hits in the event (one combination for the class $\sigma_p$,
three combinations for $\sigma_n$, and one or three for $\sigma_{\rm incl}$). 
Here $m_{\eta}$ = 547.85~MeV is the nominal $\eta$-mass, $m( 2\gamma)$ the invariant
mass of the pair of neutral hits, and $\Delta m( 2\gamma)$ its uncertainty calculated 
from the known energy and angular resolution of TAPS and CB.  
Next the same procedure was applied for the hypothesis of a $\pi^0\rightarrow 2\gamma$ 
decay (i.e. in Eq.~(\ref{eq:chi}) $m_{\eta}$ was replaced by $m_{\pi^0} =$ 134.98~MeV).
Events for which the minimum $\chi^2$ corresponded to the $\pi^0$ hypothesis were discarded
as probable background from $\pi^0$ production. 
For the other events the neutral hits corresponding to the minimum  $\chi^2$ calculated 
for $\eta$-mass were assigned to the $\eta$-decay photons. For events with an odd number of neutral 
hits (in class $\sigma_n$ or $\sigma_{\rm incl}$) the remaining neutral candidate was assigned as a
neutron candidate. In Fig. \ref{fig:confl} (top left) the confidence level distributions 
are summarized for events selected as $n\eta\rightarrow n 2\gamma$ candidates (three neutral 
hits). Shown is the distribution for the best combination from each event, the distribution
for the other two (rejected) combinations, and the distribution of the best combination
for accepted events after all further cuts. Note that the integral over the distribution of 
rejected combinations is twice the integral over the accepted ones. Most entries for the 
rejected combinations are in the first bin at a confidence level $<$~0.01 and the distribution
for accepted events is rather flat.

\begin{figure}[thb]
\resizebox{0.50\textwidth}{!}{%
  \includegraphics{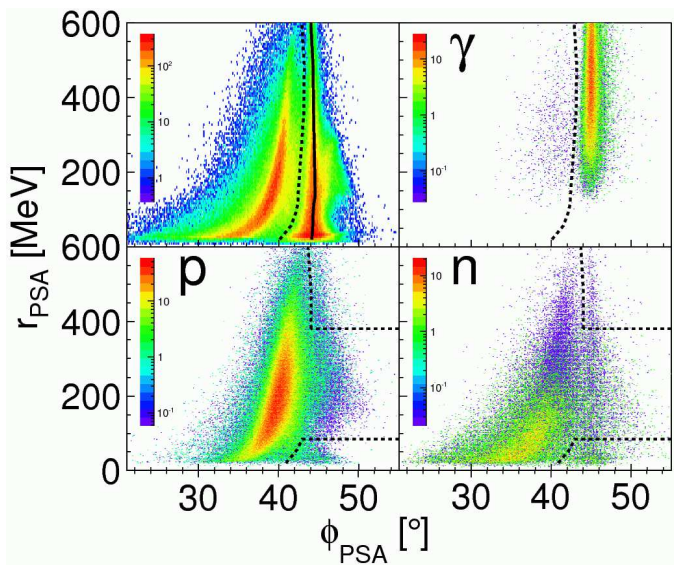}  
}
\caption{Results of TAPS pulse-shape analysis (PSA) for $\eta\rightarrow 2\gamma$; PSA radius versus 
angle for all events having passed PID, CPV cuts and $\chi^2$-analysis. Top left: all hits 
(for a typical detector module), top right: all hits assigned as photon candidates, 
bottom left: all hits assigned as proton candidates, 
bottom right: all hits assigned as neutron candidates, dashed lines indicate the cut position: $3 \sigma$ from the mean position (solid line) of the photon band. 
}
\label{fig:psa}       
\end{figure}

A similar analysis was done for the $\eta\rightarrow 6\gamma$ decays. In this case in the first
step a $\chi^2$ analysis was used to identify the most probable combination of the six or seven
neutral hits to three $\pi^0$-mesons minimizing
\begin{equation}
\chi^2 = \sum_{k=1}^{3}\frac{(m_{k}( 2\gamma)-m_{\pi^0})^2}{(\Delta
m_{k}( 2\gamma))^2}
\label{chi2}
\end{equation}
for all possible combinations of neutral pairs. Again for odd numbers of neutral hits the
left-over hit was taken as a neutron candidate. The corresponding confidence level distributions
for $n\eta\rightarrow n6\gamma$ and $p\eta\rightarrow p6\gamma$ final states are summarized
in Fig.~\ref{fig:confl} (bottom row). In this case separation for the accepted and
rejected combinations is less good for the seven-neutral hit events, simply due to combinatorics.
Events were only accepted when the invariant mass of all three neutral-hit pairs of the `best' 
combination was between 107 MeV and 163 MeV. For such events the nominal mass of the pion 
was used to improve the experimental resolution before the six-photon invariant mass
was constructed. Since the angular resolution of the detector is much better than the energy 
resolution, this was done by recalculating the photon energies and momenta from the 
approximation: 
\begin{equation}
E'_{1,2}=E_{1,2}\frac{m_{\pi^0}}{m( 2\gamma )}\;,
\label{eq:xcor}
\end{equation}
where $E_{1,2}$ are the  measured photon energies, $E'_{1,2}$ the recalculated
energies, and $m( 2\gamma)$ the measured invariant mass. 

\begin{figure}[thb]
\resizebox{0.46\textwidth}{!}{%
  \includegraphics{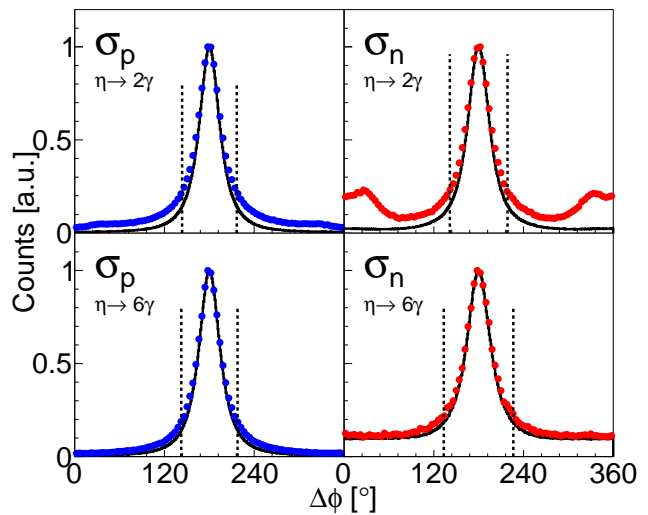}  
}
\caption{Spectra of the azimuthal angular difference between the three momenta of the recoil nucleon 
and the $\eta$-meson (`co-planarity'). Dotted lines indicate cuts. Solid curves: results from
Monte Carlo simulation with Geant4 \cite{GEANT4}.
}
\label{fig:copla}       
\end{figure}

For $\eta\rightarrow  2\gamma$ decays the photon energies were recal\-cula\-ted
in an analogous way, using the nominal $\eta$ mass, after the events had passed the
$\eta$ invariant-mass cut. 

\begin{figure*}[thb]
\centerline{\resizebox{\textwidth}{!}{%
  \includegraphics{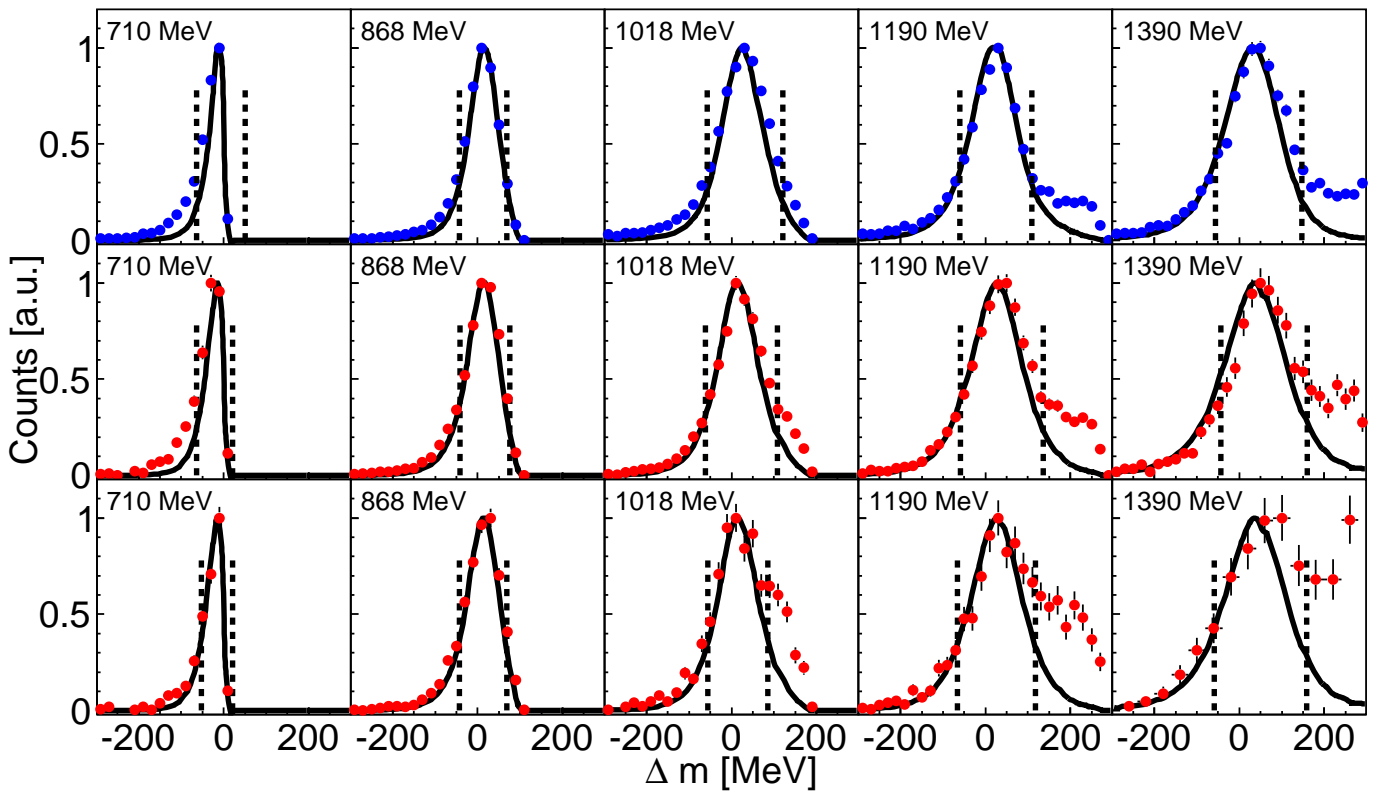} 
}}  
\caption{Missing-mass distributions for different incident photon energy ranges and the range
of invariant mass shown in Fig. \ref{fig:minv}. Closed circles are data, solid lines Geant4 
simulations. From top to bottom: $p\eta\rightarrow p 2\gamma$, $n\eta\rightarrow n 2\gamma$, and  
$n\eta\rightarrow n 6\gamma$. The cuts indicated by the dashed lines were applied to the 
invariant-mass spectra.
}
\label{fig:misma}       
\end{figure*}

In the next step the pulse-shape analysis for the TAPS detector was applied for both 
$\eta$-decay channels. It is based on the different lineshape for photon and hadron hits
in BaF$_2$ scintillators, exploited via an integration of the signals over a short (50 ns)
and a long (2 $\mu$s) time gate (see e.g. \cite{Zehr_12}). The short-gate ($E_s$) and 
long-gate ($E_l$) signal correlation was parametrized in polar coordinates $R_{\rm PSA}$, 
$\phi_{\rm PSA}$ using
\begin{equation}
R_{\rm PSA}  = \sqrt{E^2_l+E^2_s},\;\;\;\;
\phi_{\rm PSA}  =  \tan^{-1}\left(\frac{E_{s}}{E_{l}}\right)\;.
\end{equation}
Figure \ref{fig:psa} summarizes the results of the PSA. At the left-hand side, on top a spectrum
for a typical detector module with all hits is shown (the analysis was done individually for 
each module since it depends on details of the calibration). The other three plots of the figure
show for all detector modules the hits assigned by the previous analysis steps (i.e. the response 
from the charged-particle detectors and $\chi^2$ analysis) as candidates for protons, 
for neutrons, and for photons. For this analysis both detector signals, the short gate and
long gate integration, were calibrated for photon energies, so that photon hits appear
in a band at 45$^{\circ}$ ($E_{s}=E_{l}$). The photon identification is very clean and only a 
small background component at the left side of the dashed line was removed by a cut. 
The faint, residual photon band visible in the spectrum of the neutron candidates (Fig.~\ref{fig:psa})
vanishes almost after coplanarity and missing mass cuts (see below).
Since the PSA analysis cannot be reliably modeled by the Monte Carlo simulation of the
detection efficiency, and further cuts for particle identification are possible,  
very conservative cuts were made so that no significant amount of `true' recoil 
nucleons, which leaked into the photon region, was removed. Therefore, hits in the 
region of the photon band were also accepted, apart from the two regions limited by the dotted 
lines. The cuts in the upper right corner of the proton and neutron spectra are just convenient. 
They remove background from high energy photons already in an early state of the analysis, but final 
results with and without this cut are basically identical. The cut in the lower right corners is more 
important. It removes electromagnetic background (mainly electrons which have not been identified by 
the CPV). If not removed in the PSA analysis this background is cleary visible as a structure in 
the ToF-versus-E spectra, but there it is not well separated from signal events. 

The next analysis step could be done only for the event classes `$\sigma_p$' and `$\sigma_n$', 
for which the recoil nucleon was detected. The recoil nucleon and the $\eta$-meson must be 
coplanar, this means that the azimuthal angles of their three momenta must differ by 
180$^{\circ}$ (in the cm as well as in the laboratory frame as long as nuclear Fermi motion is 
neglected). 
The nucleon was assumed at rest in the initial state, which due to Fermi motion is only 
an approximation that broadens the distributions. Typical spectra for the $\sigma_p$ and 
$\sigma_n$ event classes are compared in Fig. \ref{fig:copla} to the results of Monte Carlo
simulations of the reactions with the Geant4 \cite{GEANT4} code. The background level for recoil 
protons was very low. For recoil neutrons, the background situation was different for the two- and 
six-photon decays. For the latter some combinatorial background reproduced by the Monte Carlo
simulation appears. It originates from events where the neutron was interchanged with a photon.
For the two-photon decays, an additional background component stems from the 
$\gamma n\rightarrow n\pi^0$ reaction also for the case when the neutron was mixed up with 
a photon. Accepted were only events within $\pm 2\sigma$ of the approximately Gaussian peaks
(this cut was dependent on the incident photon energy as resolution is an energy-dependent 
function). 

Subsequently the `missing mass' of all accepted events was analyzed. The recoil nucleons were 
treated as missing particles for this analysis, no matter whether they had been detected or not. 
The missing mass $\Delta m$ was constructed from the four-vector of the meson and the 
incident photon energy, again neglecting the momentum distributions of the initial-state nucleons, 
which contributes to the width of the missing-mass peaks that are centered around zero. 
The missing mass was defined by 
\begin{equation}
\Delta m = \left|{P_{\gamma}+P_{N}-P_{\eta}}\right|
-m_N\ ,
\label{eq:misma}
\end{equation}
where $m_N$ is the nucleon mass, $P_{\gamma}$, $P_{N}$, 
$P_{\eta}$ are the four-momenta of the incident photon, the initial state nucleon 
(assumed to be at rest), and the produced $\eta$-meson. Typical missing mass spectra for the 
$\sigma_p$ and $\sigma_n$ event classes from the two decay modes of the $\eta$-meson are 
summarized in Fig.~\ref{fig:misma}. Events within 1.5$\sigma$ (cut dependent on incident
photon energy) around the peak maxima were accepted for further analysis.

\begin{figure}[h]
\centerline{\resizebox{0.48\textwidth}{!}{%
  \includegraphics{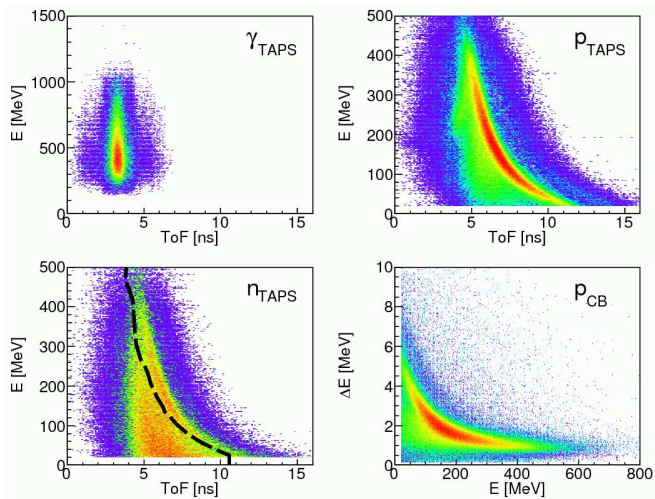} 
}}
\caption{Top, left-hand side: time-of-flight versus energy for photon candidates in TAPS;
top, right-hand side: same for proton candidates. Bottom, left-hand side: same for neutron 
candidates (events above the dashed line were rejected). Bottom,  right-hand side: $\Delta E-E$ 
for proton candidates in CB. 
}
\label{fig:taps_iden}       
\end{figure}

For those events that had passed all previous analysis steps, further redundant particle
identification signatures could be checked. For proton candidates in the CB, 
$\Delta E-E$ spectra constructed from the energy deposition in the CB and the PID were analyzed
in view of contaminations from charged pions (see Fig.~\ref{fig:taps_iden}, bottom right). 
There was, however, 
no trace of such background visible (compare Fig. 1 in Ref. \cite{Yasser_13} for a typical spectrum
with proton and charged pion band). For hits in TAPS, the ToF-versus-$E$ spectra were analyzed. 
Typical distributions are shown in Fig.~\ref{fig:taps_iden}. The spectra for photon and proton 
candidates were very clean without any trace of background. In the spectra for neutrons a small 
residual background component from misidentified protons may be visible (the TAPS charge-particle 
detector had an efficiency of better than 95\% for protons). Therefore, hits in the region of 
the proton band were rejected. 
\begin{figure}[h]
\centerline{\resizebox{0.48\textwidth}{!}{%
  \includegraphics{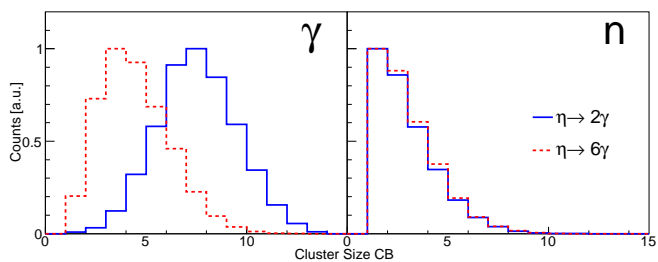}     
}}
\caption{Cluster size distributions in CB. Left-hand side: Distributions for photons, 
right-hand side: same for neutrons. Solid (blue) curves: two-photon decay of $\eta$-mesons,
dashed (red) curves: six-photon decays.
}
\label{fig:clusters}       
\end{figure}

The only signature distinguishing photon and neutron hits in the CB is the size (number of 
responding detector modules) of their clusters. The electromagnetic showers from photons spread 
over a larger number of detector modules than the energy deposition from neutrons. 
The difference 
does not allow an event-by-event separation of photon and neutron clusters. 
However, it can serve as a control of the neutron/photon assignment of hits by the methods 
discussed above. Cluster size distributions for those hits in the CB that have passed the overall 
analysis as photons or neutrons are shown in Fig.~\ref{fig:clusters}. The photon cluster sizes
differ for the two- and six-photon decays of the $\eta$-meson (because for the latter the photon
energies are smaller on average). Both photon distributions peak at higher multiplicities
than the corresponding neutron distributions, and most important, the neutron distributions
are identical for both decay channels, indicating clean photon-neutron separation. 

\begin{figure*}[thb]
\centerline{\resizebox{\textwidth}{!}{%
  \includegraphics{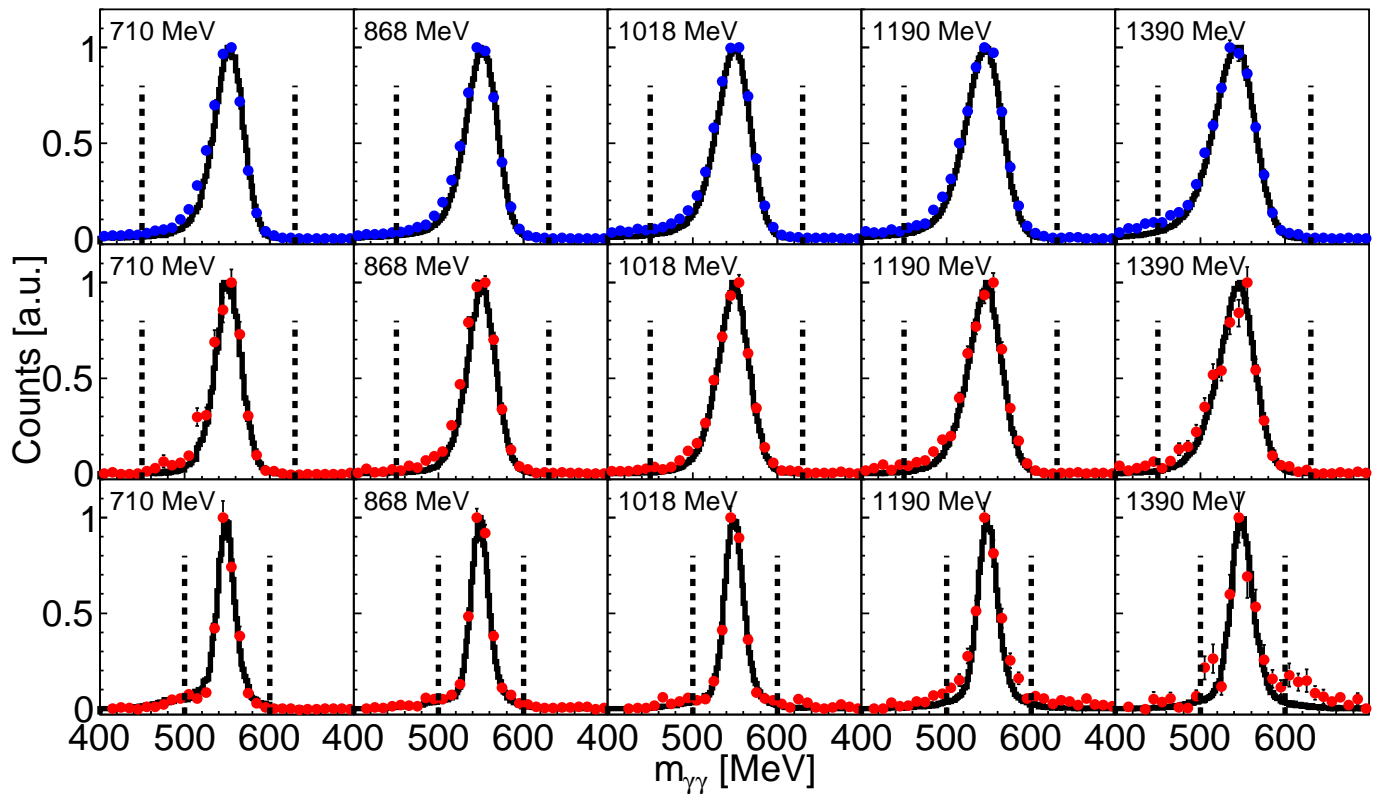}
}}
\caption{Invariant-mass distributions for different ranges of incident photon energy. 
Notation as in Fig.~\ref{fig:misma}. From top to bottom: 
$p\eta\rightarrow p 2\gamma$, $n\eta\rightarrow n 2\gamma$, and  
$n\eta\rightarrow n 6\gamma$. 
}
\label{fig:minv}       
\end{figure*}

In the final analysis step, the invariant mass of the mesons was constructed from their decay photons.
Typical spectra are summarized in Fig.~\ref{fig:minv} for the 
$\eta\rightarrow 2\gamma$ and $\eta\rightarrow 6\gamma$ decays. The spectra for the exclusive 
$\sigma_p$ and $\sigma_n$ event classes were basically free of background. Events for the two-photon 
decay were accepted for invariant masses between 450 MeV and 630 MeV and for the six-photon decay 
between 500 MeV and 600 MeV. For the inclusive data sample without recoil nucleon coincidence, a 
small residual background component was visible in the spectra, which was fitted with the simulated 
lineshape and a polynomial of degree three. 

The examples shown in Figs.~\ref{fig:misma},~\ref{fig:minv} for missing-mass and invariant-mass 
spectra were integrated over the full polar angle of the $\eta$-angular distributions. The actual 
analysis was of course done separately for each incident photon energy bin and each meson polar 
angle bin. 

Events from coherent photoproduction of $\eta$-mesons off $^3$He nuclei were not
suppressed in this analysis (they might contribute to $\sigma_{\rm incl}$). However, they are only 
significant at the lowest incident photon energies (below 650 MeV) \cite{Pheron_12} and their total 
cross section is always below 0.25~$\mu$b. Background from the target windows ($\approx$ 5\% - 10\%)
was determined with an empty target measurement and subtracted.

\subsection{Absolute normalization of cross sections and systematic uncertainties}
\label{ssub:norma}
The absolute normalization of the cross sections follows from the target surface density of 
0.073 nuclei/barn, the photon flux on the target, the detection efficiency of the calorimeter
for the different event classes, and the branching ratios for the $\eta$-decay modes. 
The uncertainty of the target density was estimated as 7\% - 8\% (measurement of target temperature
and possible deformation of the cold target cell). The uncertainty of the decay branching ratios
($b_{2\gamma}$=(39.31$\pm$0.20)\%, $b_{3\pi^0}$=(32.57$\pm$0.23)\% \cite{PDG}) is below the 1\% level
and almost negligible. The photon flux was extracted from the measurements of the flux of scattered 
electrons in the tagger focal plane and the tagging efficiency $\epsilon_{\rm tag}$, i.e. the number 
of correlated photons that pass through the collimator. The latter was measured in special low-intensity 
runs with a lead-glass detector in the photon beam (see \cite{Schumann_10} for details) and ranged from
60\% to 75\%. The resulting systematic uncertainty of the photon flux was below 5\%.
In total a systematic normalization uncertainty of 10\% was estimated in \cite{Pheron_12}.

The detection efficiency was determined with Monte Carlo (MC) simulations using the Geant4 program package
\cite{GEANT4}. It was constructed as a function of incident photon energy (respectively final-state 
invariant mass $W$) and cm polar angle of the $\eta$-mesons. The event generator for the MC simulations 
was based on the elementary cross sections for $\eta$-photoproduction off the free proton and off the 
neutron \cite{Werthmueller_13}, taking into account the momentum distribution of nucleons bound in $^3$He 
\cite{Arrington_12} (see Sec. \ref{ssub:recon}). The Geant4 program is very precise for photon showers 
but has only a limited accuracy for low-energy recoil nucleons. The detection efficiency determination 
was therefore improved with the results from measurements with the same detector configuration and 
a liquid hydrogen target. The reaction $\gamma p\rightarrow p\eta$ was analyzed in coincidence 
with recoil protons. The ratio of the numbers of $\eta$-mesons observed in coincidence with recoil 
protons to the total number of detected $\eta$ mesons corresponds to the detection efficiency for 
recoil protons. The proton detection efficiency was extracted from the measured data and compared
to the detection efficiency generated by a MC simulation of the reaction on free protons. 
The deviation between the two efficiency curves was then applied as a correction to the simulated 
detection efficiencies for quasi-free $\eta$ production off protons from the $^3$He target. 
In an analogous way, the reaction $\gamma p\rightarrow n\pi^0\pi^+$ was used 
to improve the simulation of the neutron detection efficiency. In this way, reliable detection 
efficiencies could be constructed for the full solid angle except for a small subclass of events 
for which the recoil nucleon was emitted into laboratory polar angles between 18$^{\circ}$ and 
24$^{\circ}$ (i.e. it was detected at the very edge of TAPS or the CB and had to pass through 
the support structures of CB). These data, which in the angular distributions as function of the
$\eta$ polar angle are smeared out by Fermi motion over a larger range, were interpolated in
the angular distributions as function of nucleon laboratory polar angle. The total detection efficiencies, 
depending on incident photon energy and $\eta$ polar angle, were between (first value for two-photon 
decays, values in brackets for six-photon decays) 30\% (10\%) and 80\% (50\%) for $\sigma_{\rm incl}$, 
between  4\% (4\%) and 40\% (18\%) for $\sigma_p$, and between 1\% (1\%) and 23\% (7\%) for $\sigma_n$. 
The correction factors for the detection of recoil protons derived from experiment were close to unity 
for most kinematics, but reached 20\% for low energetic protons. Correction factors for neutrons 
were also in the range between 0.8 to 1.2. 

The systematic uncertainty arising from the different analysis steps and the simulation of the detection
efficiency was estimated by a variation of all important factors. The detection efficiency of the recoil 
nucleons was additionally tested by the comparison of the inclusive cross section to the sum of 
cross sections with coincident protons and coincident neutrons. The systematic uncertainty
excluding the overall 10\% normalization uncertainty is shown by shaded bands in the figures of 
Sec.~\ref{sec:results}. 

\subsection{Kinematic reconstruction of the final state invariant mass} 
\label{ssub:recon}

Total cross-section data and angular distributions measured from bound nucleons as function of incident 
photon energy differ from the results for free nucleons due to the effects of nuclear Fermi motion.
For slowly varying cross sections this is not a large problem because the folding with the nucleon 
momentum distribution results only in a moderate energy and angular smearing of the data.
However, the effects can be significant in the vicinity of thresholds, for steep slopes, and narrow 
structures in the excitation function. They can be eliminated when the invariant mass $W$ of the 
participant nucleon and the produced meson are reconstructed from the measured four-vectors of nucleon 
and meson. The kinetic energy of recoil neutrons cannot be extracted from the deposited energy in the 
calorimeter, which is more or less random  (cf. Fig.~\ref{fig:taps_iden}). 
The kinetic energy of neutrons emitted into the solid angle covered by TAPS can be deduced from their 
time-of-flight (for neutrons in the CB the flight path is too short for reasonable resolution). 
In the energy range of interest, neutrons (or protons) detected in TAPS correspond to 
$\eta$-mesons emitted at backward angles in the photon-nucleon cm system (cos($\Theta_{\eta}^{\star}) < -0.4$). 
Only this sub-sample of data allows a direct reconstruction of the final-state invariant mass (which was done 
identically for recoil protons and neutrons).    
\begin{figure}[thb]
\centerline{
\resizebox{0.50\textwidth}{!}{%
  \includegraphics{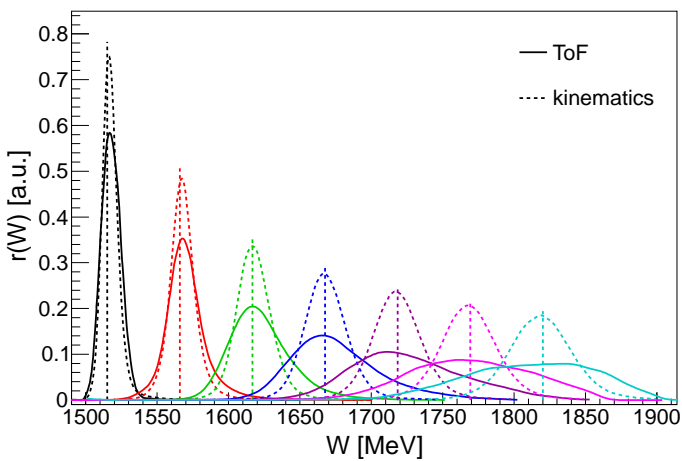}
}}
\caption{Simulated detector resolution for the invariant mass $W$ of the meson and participant nucleon 
final state reconstructed from time-of-flight measurements of the kinetic energies of the recoil nucleons 
(solid curves) and from kinematic reconstruction (dashed curves). Simulated are fixed values of $W$ 
(indicated by dashed lines) and the curves represent the detector response. 
}
\label{fig:resol}       
\end{figure}
For backward going $\eta$-mesons most often
(in particular for the two-photon decay) the photons are detected in the CB, so that there is no time 
reference signal from TAPS itself and the ToF of the recoil nucleons had to be measured with 
respect to the Tagger, which limited the time resolution (see Fig.~\ref{fig:coinc}). The achievable 
resolution was estimated by a Monte Carlo simulation of the detector response to fixed $W$ values 
(see Fig.~\ref{fig:resol}). The resolution decreases with increasing $W$ because the ToF-dependence 
on the recoil nucleon energy is rather flat for fast nucleons and the time-of-flight path was only 1.46~m.

\begin{figure}[htb]
\centerline{
\resizebox{0.50\textwidth}{!}{%
  \includegraphics{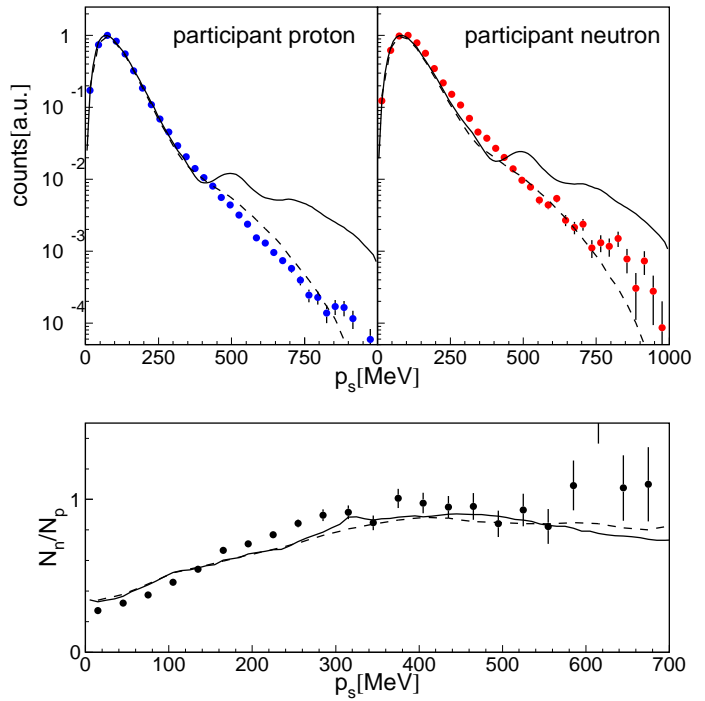}
}}
\caption{Upper part: distribution of the missing momentum of the quasi-di-nucleon spectator for participant 
protons and participant neutrons. Filled circles: present data,
solid curves: model calculation with Argonne potential for protons
(left-hand side) and neutrons (right-hand side) \cite{Arrington_12}. Dashed curves: Monte Carlo simulation
based on theory results (see text). 
Data and model normalized to unity in peak maximum. Bottom: ratio of 
neutron/proton distributions. Ratio of integrals of data normalized to $N/Z$ = 1/2 nucleon ratio of $^3$He. 
Solid (dashed) lines: model results \cite{Arrington_12} as in the upper part.
}
\label{fig:fermi}       
\end{figure}

The kinematics of quasi-free photoproduction of mesons off the deuteron can be completely reconstructed
when in addition to the four-momenta of the mesons the direction of the recoil nucleon is known 
\cite{Jaegle_11}. In this case, the momentum vector of the spectator nucleon (three variables) and the 
kinetic energy of the participant nucleon are not measured. But since the initial state (photon of known 
energy and deuteron at rest) is completely determined, these four variables can be reconstructed from 
the four equations following from momentum and energy conservation.    
For the $^3$He target, such a reconstruction can be done only in an approximate way because the two
spectator nucleons have an undetermined relative momentum $q_{\rm ps}$ in the final state. For participant 
neutrons the two spectator protons cannot be bound and for participant protons there will be a 
mixture of deuterons and unbound neutron-proton spectator pairs in the final state. However, the
relative momenta between the spectator nucleons are not large, peaking around $q$ = 70 MeV, which corresponds
to kinetic energies between 2 MeV and 3 MeV. The final state of the reaction was therefore approximated by 
the assumption that the two spectator nucleons can be treated as a di-nucleon without relative momentum.
This simplification will of course lead to a less good $W$-resolution for the $^3$He target than for the 
deuteron target. The simulated resolution under the approximation of $q_{\rm ps}=0$ is shown by the dashed 
curves in Fig.~\ref{fig:resol}; the true response will be smeared by the $q_{\rm ps}$ distribution.   

The validity of the above approximation can be tested by an analysis of the reconstructed final-state
momentum distributions of the spectator di-nucleons. As long as the $q_{\rm ps}=$ 0 approximation is reasonable
they should simply reflect the initial-state momentum distributions of the participant nucleons.
The extracted spectator momentum distributions for participant protons and neutrons are compared
in Fig.~\ref{fig:fermi} to the results from a model calculation \cite{Arrington_12} using the
Argonne potential \cite{Nogga_03}. Measured and predicted momentum distributions agree quite well for
momenta below 400 MeV. For large momenta one must take into account that (depending on the relative
orientation of the nucleon momentum with respect to the direction of the incident photon) $\eta$-production
may be kinematically forbidden. This has been investigated with a Monte Carlo simulation. The dashed
curves in the figure correspond to the results of the simulation, which used the theory curves as input 
for the momentum distributions and accepted only those events for which $\eta$-production was kinematically
allowed. The agreement with the data is quite good. Altogether the comparison suggests that the kinematic
reconstruction works well. The data reflect the main features of the momentum distribution of 
nucleons in $^3$He nuclei. The distributions peak around 70 MeV. Their ratio increases as function of the 
momentum $p_s$. At small values ($p_s <$ 300 MeV) the average ratio reflects the $N/Z$ = 1/2 ratio, while 
at larger momenta, which are mainly generated by isosinglet pairs \cite{Arrington_12}, the ratio approaches 
unity.  

\begin{figure}[thb]
\resizebox{0.50\textwidth}{!}{%
  \includegraphics{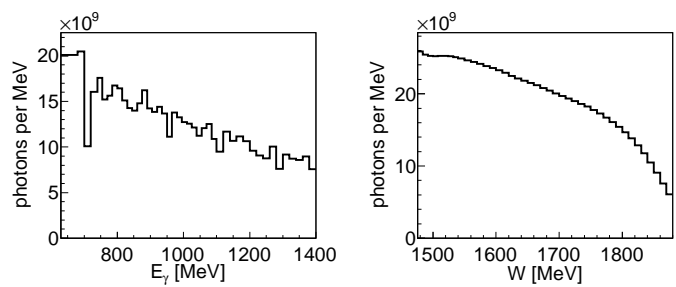}
}
\caption{Flux of incident photons. Left-hand side: measured photon flux as function of incident photon energy
$E_{\gamma}$. Right-hand side: photon flux folded with nucleon momentum distribution as function of $W$.  
}
\label{fig:flux}       
\end{figure}

For incident photon energies not too close to the $\eta$-production threshold, the reaction is dominated 
by the momentum peak close to 70 MeV (in the immediate neighborhood of the threshold larger momenta contribute 
more significantly). The average neutron-proton ratio over the momentum range up to 300 MeV, weighted with 
the probability of the Fermi momenta, is already close to the N/Z ratio of 0.5. Systematic effects of the 
momentum distributions on the magnitude of the measured neutron/proton ratio for $\eta$-production should 
therefore be small (which is supported by the comparison of the ratio to the results from the deuteron target 
in Sec.~\ref{ssec:resrecon}).   

The analysis of the kinematically reconstructed events requires also a folding of the photon flux, measured 
with the tagging spectrometer as a function of the incident photon energy, with the momentum distribution of 
the bound nucleons. This generates the effective photon flux as a function of final-state invariant mass $W$. 
The folding was done with Monte Carlo methods using the nucleon momentum distributions from 
\cite{Arrington_12} as input. The original and the folded flux distributions are shown in Fig.~\ref{fig:flux}.

\section{Results}
\label{sec:results}

In the following, all results for angular distributions are given in the cm-frame of the incident 
photon and the participant nucleon. For the Fermi-smeared results dependent on the incident photon 
energy $E_{\gamma}$, the cm-frame was constructed under the assumption that the initial state nucleon 
was at rest. For the kinematically reconstructed results as a function of $W$, the cm-frame was
derived event-by-event from the incident photon energy and the reconstructed nucleon momentum.
Total cross sections were obtained by integrating Legendre fits of the angular distributions
(see below). 

\subsection{Cross sections as function of incident photon energy}
\begin{figure*}[htb]
\resizebox{0.98\textwidth}{!}{%
  \includegraphics{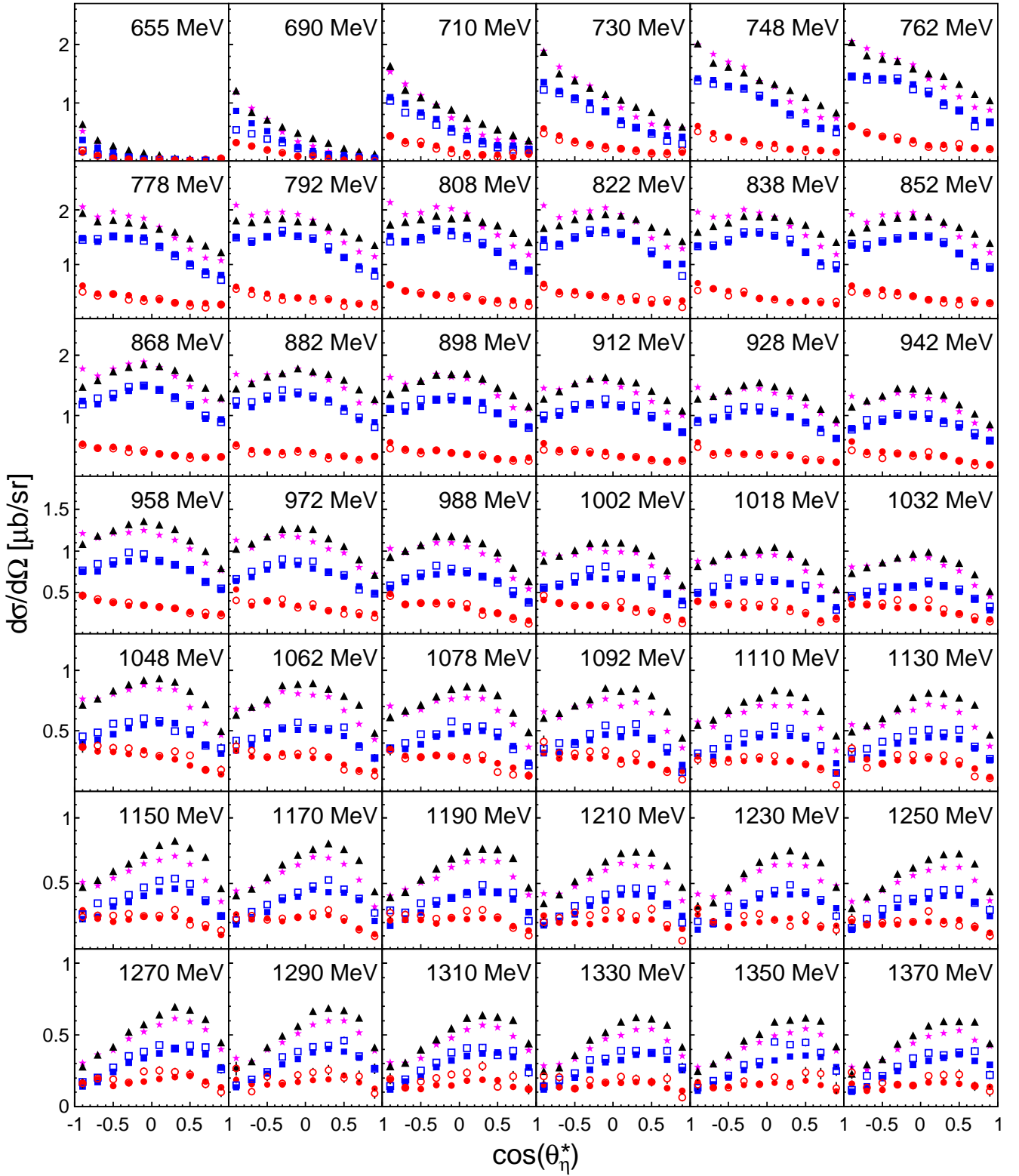}
}
\caption{Angular distributions for different bins of incident photon energy. Closed and open (red)
circles: $\sigma_n$ for two- and six-photon decay. Closed and open (blue) squares: same for
2$\sigma_p$. (Black) triangles: average of two- and six-photon decay for $\sigma_{\rm incl}$.
(Magenta) stars: same for $2\sigma_p + \sigma_n$.}
\label{fig:diff_e}       
\end{figure*}

\begin{figure}[thb]
\centerline{
\resizebox{0.47\textwidth}{!}{%
  \includegraphics{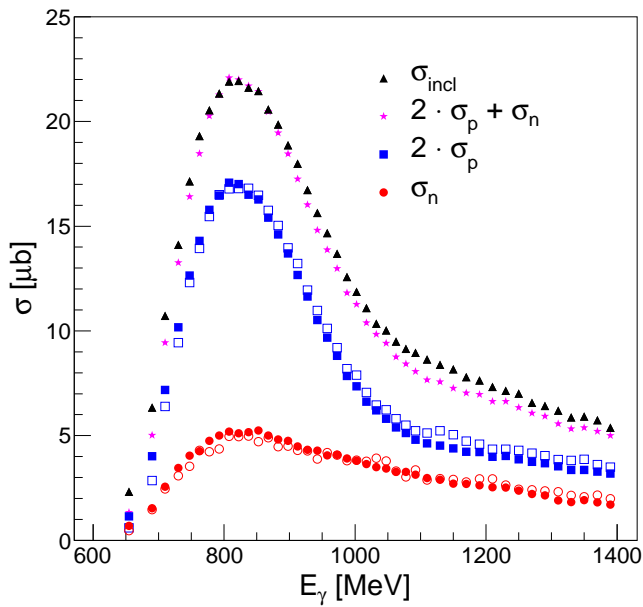}
}}
\caption{Total cross sections as a function of incident photon energy. For $\sigma_p$ and $\sigma_n$
filled symbols represent the results from the $\eta$ two-photon decay, open symbols results from
the six-photon decay. For the inclusive cross section $\sigma_{\rm incl}$ and the sum of proton and
neutron cross section only the average of both decay channels is shown.
}
\label{fig:total_e}       
\end{figure}

The cross section as function of incident photon energy was extracted for the three event classes
$\sigma_p$, $\sigma_n$, and $\sigma_{\rm incl}$ defined in Sec.~\ref{ssec:iden} independently
for two-photon and six-photon decays of the $\eta$-mesons. Angular distributions are shown in
Fig.~\ref{fig:diff_e}, the results for the total cross sections are summarized in Fig.~\ref{fig:total_e}. 
For the reactions in coincidence with protons and with neutrons, results for both $\eta$-decay 
channels are shown. The comparison allows an estimation of systematic effects from the reaction 
identification and background elimination.

The total inclusive reaction cross section $\sigma_{\rm incl}$ is compared to the sum
of proton and neutron results as an independent cross check for the used proton and neutron
detection efficiencies. For $\sigma_{\rm incl}$ only the $\eta$-detection efficiency enters, 
while $\sigma_p$ ($\sigma_n$) are in addition based on proton (neutron) detection efficiencies.
The largest deviations between $\sigma_{\rm incl}$ and $2\sigma_p + \sigma_n$ were on the 10\% level
at larger incident photon energies. Most likely, they are due to residual background in the
$\sigma_{\rm incl}$ data for which cuts based on recoil nucleon detection (co-planarity) 
are precluded. This effect is more important for higher incident photon energies, where the
background from $\eta\pi$-pairs is large.

\clearpage

A comparison of the proton and neutron data shows that, as for the deuteron target, the rise of 
the neutron excitation function below incident photon energies of 1 GeV towards the $S_{11}$(1535) 
maximum is less steep than for the proton. But as expected, in contrast to the deuteron target 
\cite{Jaegle_11}, there is no pronounced structure visible in the neutron excitation function 
around 1 GeV. Due to the larger momenta of nucleons bound in $^3$He any such structures will be 
smeared out. 

Note that for the angular distributions in Fig.~\ref{fig:diff_e}, as for the total cross sections, 
the proton results correspond to $2\sigma_p$. The comparison of the results
for the two $\eta$-decay channels and between $\sigma_{\rm incl}$ and $2\sigma_p + \sigma_n$
demonstrates again the internal consistency of the data. 
The strong rise of the angular distributions in the threshold region towards backward angles  
is only an artifact from the choice of the cm-system. The elementary cross sections from the 
free proton (and neutron) are rather flat in this energy range. However, here the reactions were
analyzed in the cm-frame of the incident photon and a nucleon at rest, neglecting its
Fermi motion. 
At higher incident photon energies this causes only a moderate smearing of the
angular distributions. But close to threshold the situation is asymmetric because only 
kinematics with antiparallel photon and nucleon momentum can give rise to $\eta$-production.
This means that on average the photon - nucleon-at-rest frame is faster in the lab system than
the `true' photon - nucleon cm-frame and the mesons appear backward boosted in the used frame.   

\begin{figure}[htb]
\resizebox{0.46\textwidth}{!}{%
  \includegraphics{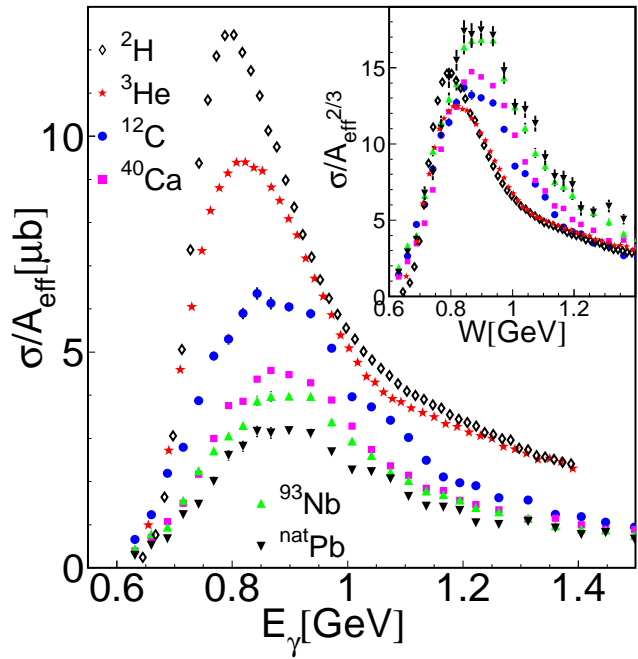}
}
\caption{Total inclusive cross section $\sigma_{\rm incl}$ compared to results from other nuclei
($^2$H from \cite{Werthmueller_13}, other nuclei from \cite{Mertens_08}.
In the main plot the cross sections are scaled by $A_{\rm eff} = N_p + (2/3) N_n$ (see text);
in the insert by $A_{\rm eff}^{2/3}$.
}
\label{fig:total_comp}       
\end{figure}

Previously, quasi-free inclusive $\eta$-photoproduction off nuclei in the $S_{11}$(1535) range 
was investigated \cite{Roebig_96,Mertens_08} in view of $\eta$-nucleus interactions and possible 
in-medium effects on the $S_{11}$ resonance. The data are compared in Fig.~\ref{fig:total_comp}
to the present results for $^3$He nuclei. They were scaled not by the mass numbers $A$ of the nuclei
but by effective mass numbers defined by $A_{\rm eff} = N_p + (2/3)N_n$ with the proton number
$N_p$ and the neutron number $N_n$. This scaling was chosen because in the range of interest (in
the $S_{11}$(1535) maximum) the elementary cross sections off free protons and off free neutrons 
are related by $\sigma_n/\sigma_p =$ 2/3. The main plot of Fig.~\ref{fig:total_comp} compares the
data scaled by $A_{\rm eff}$; in the insert they are scaled by $A_{\rm eff}^{2/3}$. The position
of the $S_{11}$ resonance maximum shifts to higher incident photon energies for the heavier
nuclei, but this is mainly a trivial effect from the shift of the momentum distribution of the
bound nucleons to higher momenta for heavier nuclei; it saturates for $^{40}$Ca.
In the energy range below the threshold for the production of $\eta\pi$-pairs (i.e. in the rising slope
of the $S_{11}$ resonance), the data scale almost perfectly with $A_{\rm eff}^{2/3}$ 
(i.e. with the nuclear surface), which confirms the strong absorption of $\eta$-mesons in nuclei. 
It is surprising that this scaling is still valid for light nuclei such as $^3$He.
Even for the deuteron data it holds true approximately, only the rising slope is a bit steeper due to
the much smaller nucleon momenta. The deviations from the $A_{\rm eff}^{2/3}$ scaling above
the $S_{11}$ peak are partly due to residual background from $\eta\pi$ final states, which becomes
significant for incident photon energies above 800 MeV (see \cite{Mertens_08}). It is more important
for the heavier nuclei because of the poorer separation of single $\eta$ and $\eta\pi$ final
states in missing-mass spectra caused by the larger Fermi momenta.

\begin{figure*}[htb]
\resizebox{1.0\textwidth}{!}{%
  \includegraphics{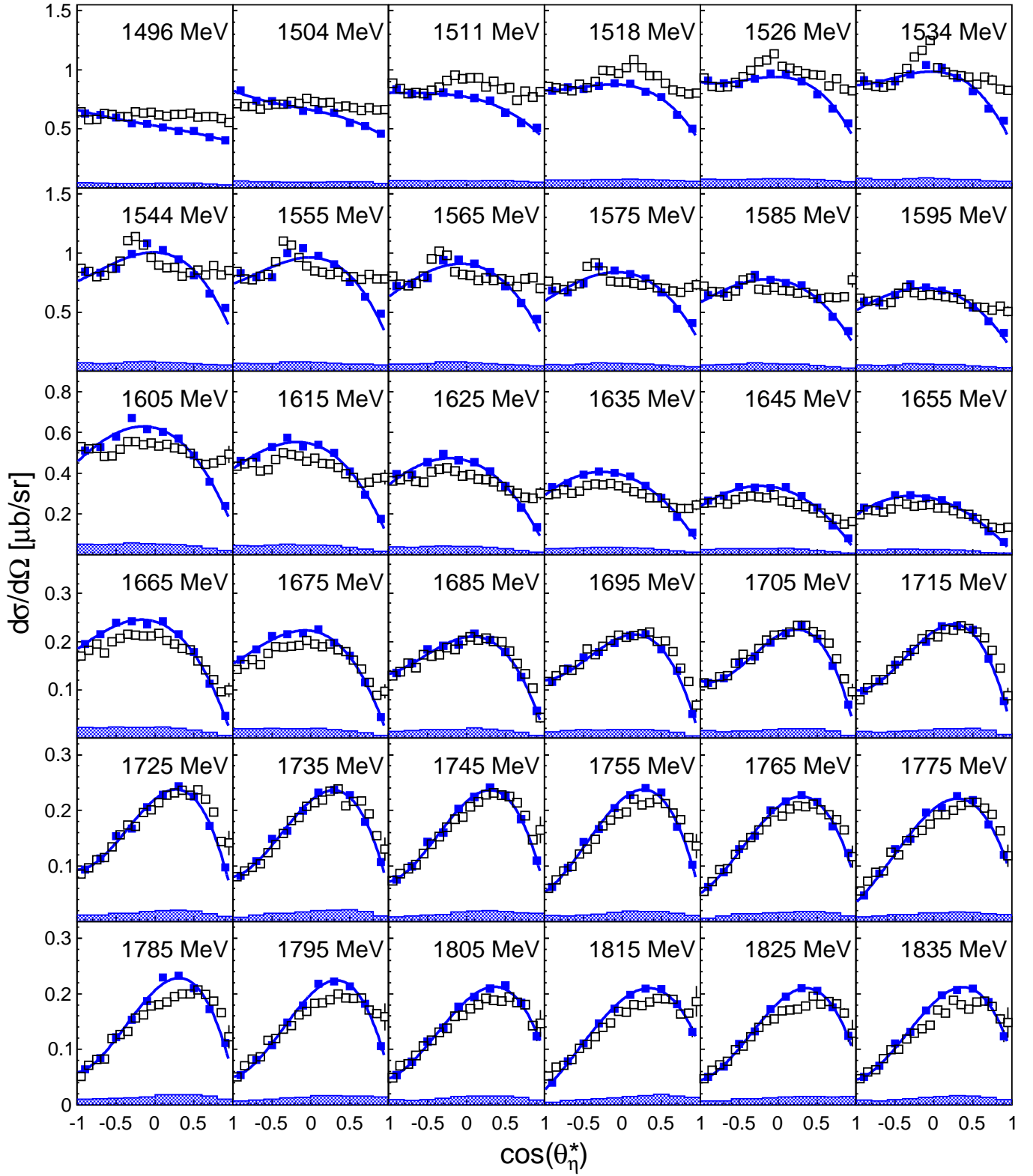}
}
\caption{Angular distributions of the quasi-free $\gamma p\rightarrow p\eta$ reaction.
(Blue) closed squares: data from present $^3$He measurement, open (black) squares: data from
deuterium target \cite{Werthmueller_13} scaled down by factor 0.75. Solid lines: fits of 
data with Legendre polynomials. Histograms at bottom of figures: systematic uncertainties 
of present data, except 10\% total normalization uncertainty.
}
\label{fig:diff_wp}       
\end{figure*}

\clearpage

\begin{figure*}[htb]
\resizebox{1.0\textwidth}{!}{%
  \includegraphics{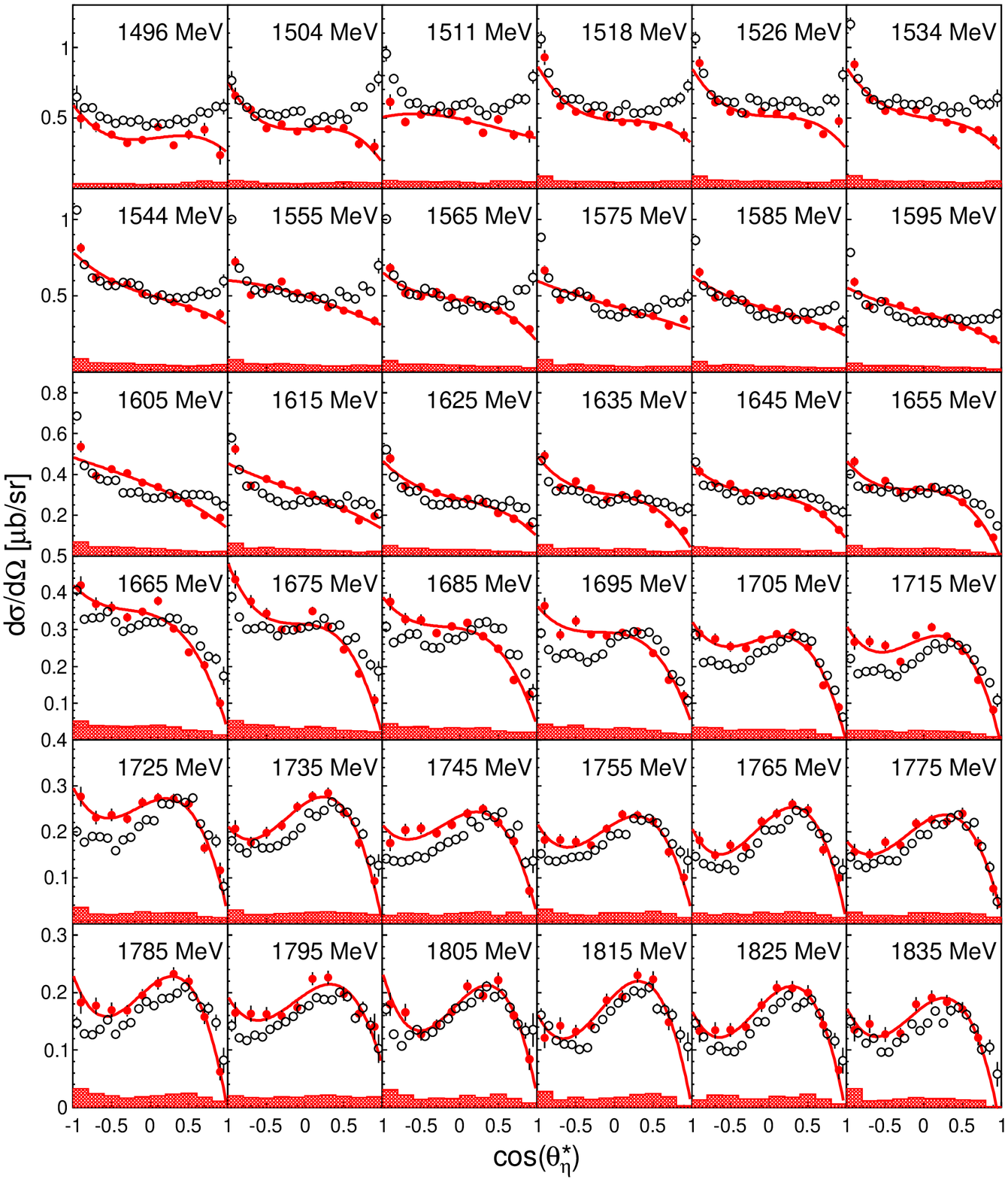}
}
\caption{Angular distributions of the quasi-free $\gamma n\rightarrow n\eta$ reaction.
(Red) closed circles: data from present $^3$He measurement, open (black) circles: data from
deuterium target \cite{Werthmueller_13} scaled down by factor of 0.75. Solid lines: fits of 
data with Legendre polynomials. Histograms at bottom of figures: systematic uncertainties 
of present data, except 10\% total normalization uncertainty.
}
\label{fig:diff_wn}       
\end{figure*}

\clearpage

\begin{figure}[htb]
\resizebox{0.49\textwidth}{!}{%
  \includegraphics{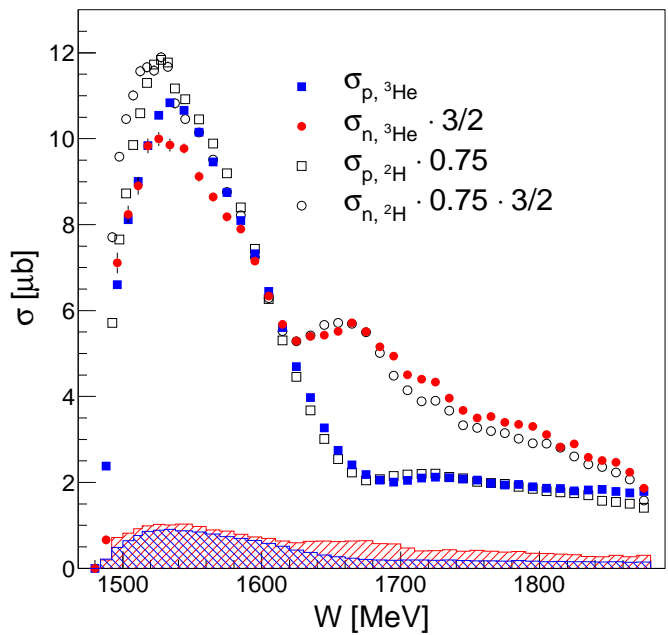}
}
\caption{Total cross sections as a function of reconstructed final-state invariant mass $W$.
(Blue) squares: proton coincidence $\sigma_p$, (red) filled circles: neutron coincidence 
$\sigma_n$. (Black) open squares and open circles: corresponding excitation functions measured
with a deuterium target \cite{Werthmueller_13}. Neutron data scaled up by a factor of 3/2,
data from deuterium target scaled down by a factor of 0.75. Shaded bands: systematic uncertainty
of $^3$He data (excluding the overall normalization uncertainty of 10\%), single (red) hatched:
neutron, double (red/blue) hatched: proton). All data averaged over two- and six-photon 
decays of the $\eta$-meson.
}
\label{fig:total_w}       
\end{figure}

\begin{figure}[thb]
\resizebox{0.50\textwidth}{!}{%
  \includegraphics{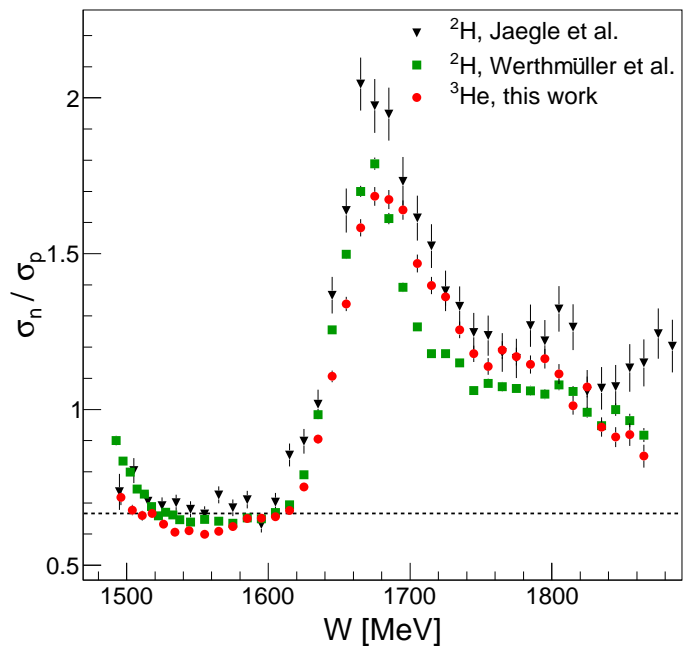}
}
\caption{Cross-section ratio $\sigma_n/\sigma_p$ for quasi-free $\eta$ production. Filled (red)
circles from present work for $^3$He target, filled (green) squares: deuterium target 
Ref. \cite{Werthmueller_13}, filled (black) triangles: deuterium target Ref. \cite{Jaegle_11}.  
}
\label{fig:ratio}       
\end{figure}

\subsection{Cross sections as a function of reconstructed final-state invariant mass}
\label{ssec:resrecon}

The angular distributions for the reaction off quasi-free protons and 
off quasi-free neutrons from the kinematic reconstruction are summarized in 
Figs.~\ref{fig:diff_wp},~\ref{fig:diff_wn}, the total cross sections are shown
in Fig.~\ref{fig:total_w}. The latter have been extracted from the integration of Legendre 
polynomials fitted to the angular distributions.

The results are compared in the figures to similar data from a deuterium target 
\cite{Werthmueller_13}. 
Comparison of the total cross sections in Fig.~\ref{fig:total_w} shows that,
apart from the immediate threshold region where effects from Fermi motion are most important,
the energy dependence of the present data is in excellent agreement with that for the deuteron data.
The absolute magnitude of the data differs between $^3$He and the deuteron by a factor of 
$\approx$ 0.75 for quasi-free protons as well as for quasi-free neutrons, independent of
incident photon energy. 

Angular distributions from the $^3$He measurement and the deuteron 
target agree reasonably well for $W$ above $\approx$1.6 GeV (apart from the 0.75 overall scaling factor).
The difference in magnitude of the cross sections is clearly beyond the range of systematic
uncertainty of the data. Here one must also take into account that both measurements were
performed with an identical detector setup and were analyzed with the same tools using similar 
cuts for particle and reaction identification and the same Monte Carlo code for the simulation
of the detection efficiency. Thus a large part of the quoted systematic uncertainty cancels
in the comparison of the results. The deviation in absolute magnitude must be due to 
different nuclear effects in quasi-free $\eta$-photoproduction off the deuteron and $^3$He
nuclei. At least part of it may arise from the approximation used in the kinematic reconstruction 
that the relative momentum between the two spectator nucleons vanishes. This will not only 
have an effect on the shape of the angular distributions (which is most apparent in the threshold 
region) but can also lead to a reduction in magnitude. This is so because the approximation was 
also made for the folding of the photon flux with the nucleon momentum distribution. 
Non-vanishing relative momenta of the spectators reduce the available energy in the 
meson-participant-nucleon system, which can suppress $\eta$ production. This effect will tend to 
overestimate the available photon flux for a given final-state invariant mass $W$.

A quantitative 
analysis of such effects is presently not available. Furthermore, FSI effects, 
either between $\eta$-meson and nucleons or among the nucleons, may be important. The 
scaling of the inclusive cross section, discussed in the previous section, suggests that such 
effects are non-negligible for $^3$He.

The most important result is that in the range above $W$ = 1.6 GeV, which is mainly 
of interest because of the observed structure in the neutron excitation function, 
good agreement was found for the data from $^3$He and $^2$H, apart from the overall scaling. 
This is even more apparent from a comparison of the measured neutron/proton cross section 
ratios in Fig.~\ref{fig:ratio}. The data measured for $^3$He and the deuteron consistently show
a rapid change of the ratio with a narrow peak around 1.67~GeV. At lower $W$ values all data sets
agree with the $\sigma_n/\sigma_p$=2/3 cross section ratio of the $S_{11}$(1535) excitation.

\begin{figure}[thb]
\resizebox{0.48\textwidth}{!}{%
  \includegraphics{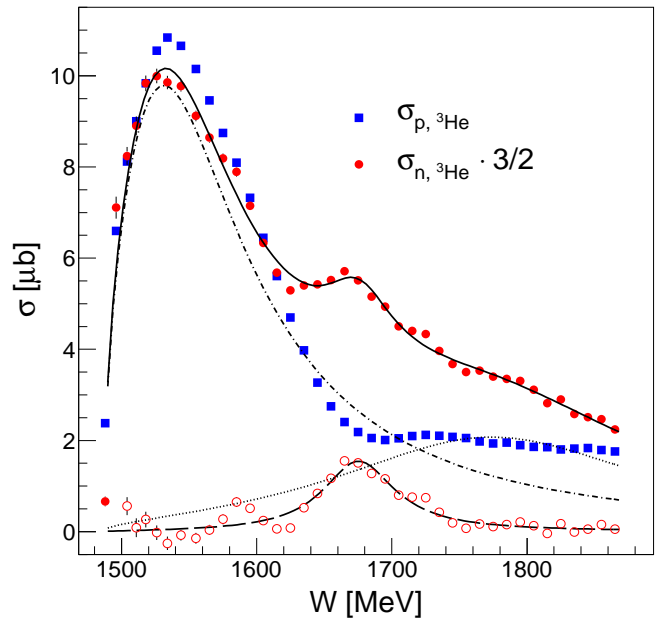}
}
\caption{Phenomenological fit of neutron excitation function. Filled (blue) squares: proton excitation
function, filled (red) circles: neutron excitation function, open red circles: neutron excitation 
function after subtraction of $S_{11}$ and background fit. Curves correspond to: fit result for
$S_{11}$(1535) Breit-Wigner curve (dash-dotted), phenomenological Breit-Wigner curve for background 
parametrization (dotted), Breit-Wigner curve for fit of narrow structure (long-dashed), and sum of all
(solid). 
}
\label{fig:total_fit}       
\end{figure}

The neutron excitation function has been fitted in the same way as the deuteron data in \cite{Jaegle_11}
by the sum of an energy dependent Breit-Wigner (BW) curve for the $S_{11}$(1535) resonance, a 
BW curve for the narrow structure and a further broad BW curve parameterizing effectively
all contributions from further resonances and non-resonant background. The result is shown
in Fig.~\ref{fig:total_fit}. The fit parameters are summarized in Table~\ref{tab:bwfit} and compared
to the results from the measurements with deuterium targets \cite{Jaegle_11,Werthmueller_13}.

\begin{figure}[h]
\resizebox{0.48\textwidth}{!}{%
  \includegraphics{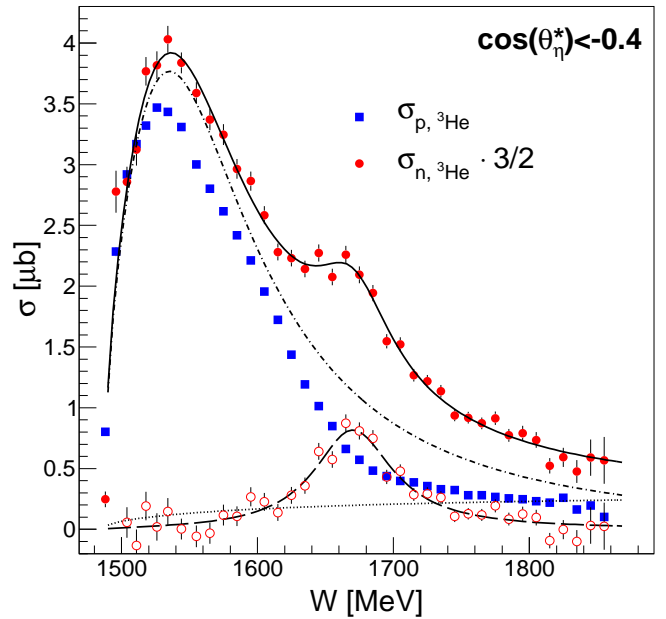}  
}
\caption{Integrated cross section for polar angles with cos($\Theta_{\eta}^{\star})<$~-0.4 as function of $W$.
Kinematics reconstructed using the ToF measurement for the recoil nucleons.
}
\label{fig:total_tof}       
\end{figure}

\begin{figure}[thb]
\resizebox{0.49\textwidth}{!}{%
  \includegraphics{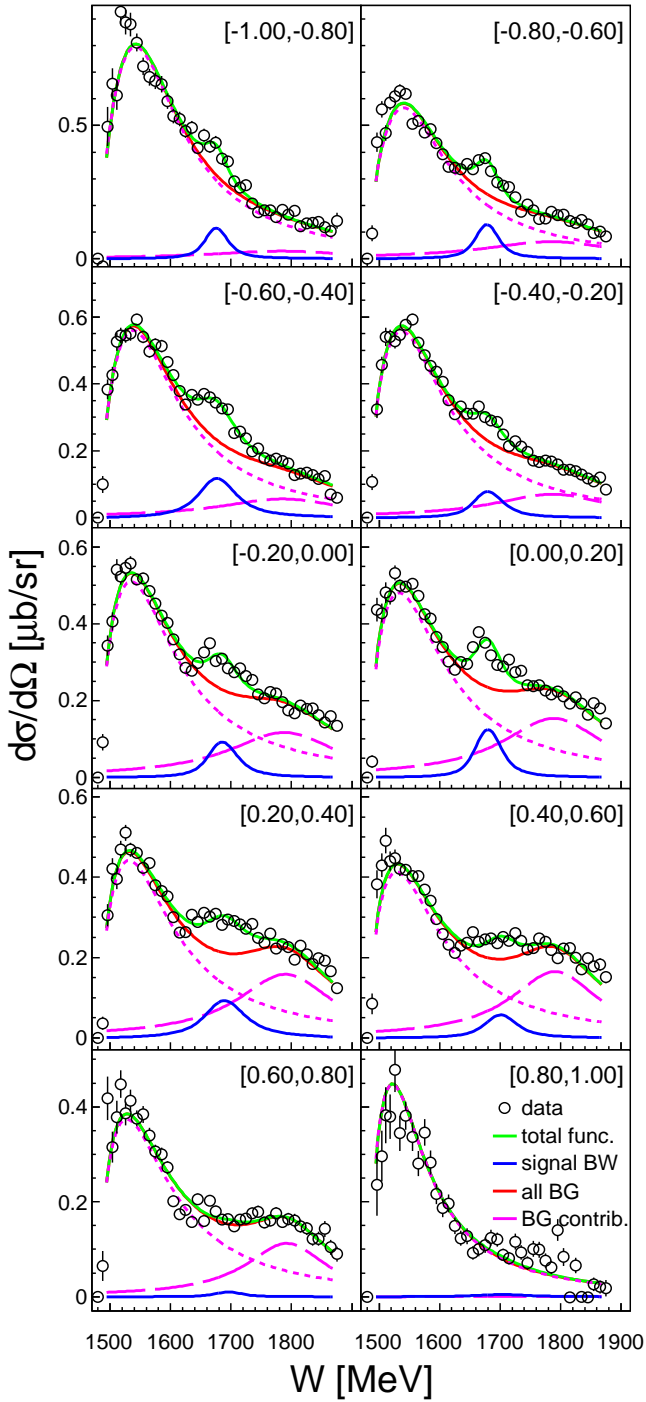} 
}
\caption{Excitation functions for $\gamma  n\rightarrow n \eta$ for different ranges of $\eta$ cm 
polar angle. The curves represent short dashed (magenta): BW-fit of S$_{11}$(1535), 
long dashed (magenta) phenomenological background, solid (red): sum of both, 
solid (blue): BW fit of narrow structure, solid (green): complete fit. 
}
\label{fig:ang_exci}       
\end{figure}
 
All results for widths given in Table~\ref{tab:bwfit} are effective quantities including the broadening 
of structures due to the experimental resolution. The quoted uncertainties are only statistical. Typical 
systematic effects in the fits can be estimated from a comparison of the results for the $S_{11}$(1535). 
Its position varies by $\approx$10~MeV ($\sim$0.65\%) and its width by $\approx$15~MeV ($\sim$10\%)
and the differences between the two deuterium measurements are comparable to those between deuterium and
$^3$He.   

\begin{table}[h]
\begin{center}
\caption{Fitted parameters of the $S_{11}$(1535) resonance  `($S_{11}$)' and the narrow structure 
`(X)' compared to the measurements with a deuterium target from Werthm\"uller et al. 
\cite{Werthmueller_13} and Jaegle et al. \cite{Jaegle_11}. $^{\dagger}$For the latter the result 
given in the paper corresponds to an analysis with a strong cut on the Fermi momentum of the 
spectator nucleon.
Without this cut the width of the narrow structure is 54$\pm$16~MeV. The present results labeled 
(`ToF') are from the analysis of the restricted angular range accessible for ToF measurements 
of the recoil nucleon.
} 
\label{tab:bwfit}       
\begin{tabular}{|c|c|c|c|}
\hline\noalign{\smallskip}
& $W_R$ [MeV] & $\Gamma$ [MeV] & $\sqrt{b_{\eta}}A_{1/2}^n$ \\
& & & [10$^{-3}$GeV$^{-1/2}$] \\
\hline
$^2$H (X) \cite{Jaegle_11}           & 1663$\pm$3 & 25$\pm$11$^{\dagger}$  & 12.2$\pm$3 \\
$^2$H ($S_{11}$) \cite{Jaegle_11}       & 1535$\pm$4 & 167$\pm$23 & 62$\pm$2       \\
 \hline
$^2$H (X) \cite{Werthmueller_13}     &  1670$\pm$1 & 50$\pm$2   &  12.3$\pm$0.8    \\ 
$^2$H ($S_{11}$) \cite{Werthmueller_13} &  1527$\pm$1 & 147$\pm$2  & 79$\pm$1      \\
\hline
$^3$He (X)                           &  1675$\pm$2  &  62$\pm$8 & 11.9$\pm$1.2     \\
$^3$He ($S_{11}$)                       &  1536$\pm$1  & 162$\pm$4 & 66$\pm$1      \\
\hline
$^3$He (X) (ToF)                     &  1671$\pm$2  &  61$\pm$10 & -                \\
$^3$He ($S_{11}$) (ToF)                 &  1541$\pm$2  & 174$\pm$10 & -             \\
\hline
\end{tabular}
\end{center}
\end{table}

The results from Ref. \cite{Werthmueller_13} for the deuterium measurement were also fitted with
a BW curve folded with the experimental resolution. The intrinsic width $\Gamma_0$ of the narrow structure
determined this way was 29$\pm$3~MeV. A rough approximation of the intrinsic width of the structure from 
the $^3$He data is $\Gamma_0\approx$ 45~MeV, obtained by just subtracting in quadrature the width 
($\approx$~35~MeV FWHM) of the corresponding resolution curve from Fig.~\ref{fig:resol}. Here one must 
keep in mind that due to the approximation (no relative momentum of spectator nucleons) made in the
simulation of the resolution, this is only an upper limit. 

In Fig.~\ref{fig:total_tof} excitation functions for quasi-free protons and neutrons over
a restricted range of meson polar angles are shown. These were constructed using the TAPS ToF measurement
to determine the recoil nucleon energies. For this kinematic reconstruction no approximations have 
to be made, since the four-momenta of $\eta$-meson and recoil nucleon are directly measured.
However, as shown in Fig.~\ref{fig:resol}, the expected resolution is rather low. The results of
the fit, done in the same way as for the total cross section, are also shown in Table \ref{tab:bwfit}.
They are similar to the parameters obtained from the fit in Fig.~\ref{fig:total_fit}
although the shape of the phenomenological background is much different for this restricted angular
range. If we take into account the experimental resolution from Fig.~\ref{fig:resol}, the effective width 
of $\approx$~60~MeV must correspond to an intrinsic width $\Gamma_0$ of the structure below the 
40~MeV range. Altogether, the results from the $^3$He and $^2$H targets are consistent with a structure
in the $\gamma n\rightarrow n\eta$ excitation function with an intrinsic width of less than 40~MeV.

Finally, excitation functions in bins of cm polar angle are summarized in Fig.~\ref{fig:ang_exci}. 
They have been fitted with the same ansatz used for the total cross section. The results show a 
behavior that agrees with the deuterium results \cite{Werthmueller_13}. A structure around 
$W$ = 1.67~GeV is visible for almost all angles but it varies in strength and shape in a 
non-trivial way across the angular distribution, which will have to be analyzed with 
partial-wave analyses.

\section{Summary and conclusions}

Precise total and differential cross-section data were measured for photoproduction of
$\eta$-mesons off quasi-free protons and neutrons bound in $^3$He nuclei. 
The narrow structure previously observed for $\eta$ photoproduction off neutrons bound in the 
deuteron is clearly observed also for the $^3$He measurement and the parameters of this effect 
- position, width, excitation strength - agree with the analysis of the deuteron data, except
for an overall difference in magnitude. The latter is observed for both quasi-free reactions,
i.e. for participant neutrons as well as for participant protons. The quasi-free reaction cross 
sections are suppressed in comparison to the deuteron target by a factor of $\approx$~0.75,
almost independent of photon energy. This effect is outside the systematic normalization
uncertainty of the data. However, the ratio of the quasi-free neutron and proton excitation 
functions agrees very well with the results from deuteron targets when the effects from nuclear 
Fermi smearing are eliminated by a kinematic reconstruction of the invariant mass of the 
participant nucleon and meson in the final state. Apart from the immediate threshold region,
good agreement between the $^3$He and deuterium data is also found for the shape of the angular
distributions. The overall scale difference can be partly related to residual effects from
the momentum distributions of the bound nucleons. Finite relative momenta between the
two `spectator' nucleons can kinematically reduce the production probability of the relatively 
heavy $\eta$-mesons. FSI effects can also play a role. The observed scaling of 
the total inclusive cross section with mass number, derived from a comparison with $\eta$-production
data for heavier nuclei, indicates that such effects are already significant for the 
helium nucleus. Modelling of the reaction in a distorted-wave impulse approximation with realistic 
$^3$He wave functions, taking into account the FSI effects would be desirable, but is not 
available as yet. A comparison of the present data, the quasi-free data from the deuteron,
and free proton data offers a good basis for the study of the nuclear effects. Independent
of this discussion, the present results demonstrate that the narrow structure observed
for $\eta$ photoproduction off neutrons bound in the deuteron is robust enough to survive
also in a much different nuclear environment and is thus almost certainly not related to
some nuclear effect but a feature of the elementary $\gamma n\rightarrow n\eta$ reaction.

\vspace*{1cm}
{\bf Acknowledgments}

We wish to acknowledge the outstanding support of the accelerator group 
and operators of MAMI. 
This work was supported by Schweizerischer Nationalfonds
(200020-132799,121781,117601,113511), Deutsche
For\-schungs\-ge\-mein\-schaft (SFB 443, SFB/TR 16), DFG-RFBR (Grant No. 05-02-04014),
UK Science and Technology Facilities Council, (STFC 57071/1, 50727/1), 
European Community-Re\-search Infrastructure Activity (FP6), the US DOE, US NSF and
NSERC (Canada).
We thank the undergraduate students of Mount Allison University and The George Washington  
University for their assistance.

\end{document}